\documentclass[12pt, draftclsnofoot, onecolumn]{IEEEtran}
\usepackage{amsmath}
\usepackage{amssymb}
\usepackage{amsfonts}

\newcommand{\Rmnum}[1]{\expandafter\@slowromancap\romannumeral #1@}
\usepackage{graphicx}
\usepackage{epsfig}
\usepackage{subfigure}
\usepackage{citesort}
\usepackage{xcolor}
\usepackage{enumerate}
\usepackage{extarrows}
\usepackage{algpseudocode, algorithm}

\makeatletter
\algrenewcommand\ALG@beginalgorithmic{\small}
\makeatother

\begin{document}
\title{Optimizing DF Cognitive Radio Networks with Full-Duplex-Enabled Energy Access Points}

\author{Hong Xing, Xin Kang, Kai-Kit Wong, and Arumugam Nallanathan
\thanks{Part of this paper has been accepted by the IEEE Global Communications Conference (GLOBECOM), 2016.}
\thanks{H. Xing and A. Nallanathan are with the Department of Informatics, King's College London, London, WC2R 2LS, UK (e-mails: $\rm hong.xing@kcl.ac.uk$; $\rm arumugam.nallanathan@kcl.ac.uk$).}
\thanks{X. Kang is with the National Key Laboratory of Science and Technology on Communications, University of Electronic Science and Technology of China, Chengdu, 611731, China (e-mail: $\rm kangxin83@gmail.com$).}
\thanks{K.-K. Wong is with the Department of Electronic and Electrical Engineering, University College London, London, WC1E 6BT, UK (e-mail: $\rm kai$-$\rm kit.wong@ucl.ac.uk$).}}
\maketitle
\begin{abstract}
With the recent advances in radio frequency (RF) energy harvesting (EH) technologies, wireless powered cooperative cognitive radio network (CCRN) has drawn an upsurge of interest for improving the spectrum utilization with incentive to motivate joint information and energy cooperation between the primary and secondary systems.
Dedicated energy beamforming (EB) is aimed for remedying the low efficiency of wireless power transfer (WPT), which nevertheless arouses out-of-band EH phases and thus low cooperation efficiency. To address this issue, in this paper, we consider a novel RF EH CCRN aided by full-duplex (FD)-enabled energy access points (EAPs) that can cooperate to wireless charge the secondary transmitter (ST) while concurrently receiving primary  transmitter (PT)'s signal in the first transmission phase, and to perform decode-and-forward (DF) relaying in the second transmission phase. We investigate a weighted sum-rate maximization problem subject to the transmitting power constraints as well as a total cost constraint using successive convex approximation (SCA) techniques. \textcolor{black}{A zero-forcing (ZF) based suboptimal scheme that requires only local CSIs for the EAPs to obtain their optimum receive beamforming is also derived.} Various tradeoffs between the weighted sum-rate and other system parameters are provided in numerical results to corroborate the effectiveness of the proposed solutions against the benchmark ones.
\end{abstract}

\IEEEpeerreviewmaketitle
\newtheorem{definition}{\underline{Definition}}[section]
\newtheorem{fact}{Fact}
\newtheorem{assumption}{Assumption}
\newtheorem{theorem}{\underline{Theorem}}[section]
\newtheorem{lemma}{\underline{Lemma}}[section]
\newtheorem{proposition}{\underline{Proposition}}[section]
\newtheorem{corollary}[proposition]{\underline{Corollary}}
\newtheorem{example}{\underline{Example}}[section]
\newtheorem{remark}{\underline{Remark}}[section]
\newcommand{\mv}[1]{\mbox{\boldmath{$ #1 $}}}
\newcommand{\Myfrac}[2]{\ensuremath{#1\mathord{\left/\right.\kern-\nulldelimiterspace}#2}}

\section{Introduction}
With the rapid development of wireless services and applications, the demand for frequency resources has dramatically increased. How to accommodate these new wireless services and applications within the limited radio spectrum becomes a big challenge facing the modern society \cite{FCC02}. The compelling need to establish more flexible spectrum regulations motivates the advent of cognitive radio (CR) \cite{Mitola1999CR}. Cooperative cognitive radio networks (CCRNs) further pave way to improve the spectrum efficiency of a CR system by advocating cooperation between the primary and secondary systems for mutual benefits. Compared with classical CR approaches \cite{kangTWC}, CCRN enables cooperative gains on top of CR in the sense that the secondary transmitter (ST) helps to provide the diversity and enhance the performance of primary transmission via relaying the primary user (PU)'s message while being allowed to access the PU's spectrum.

Although the conventional CCRN benefits from information-level cooperation, its implementation in real world might be limited due to STs' power constraints, especially when the STs are low-power devices, such as energy constrained wireless sensors and small cell relays. With the advent of various energy harvesting (EH) technologies, CCRN has now been envisioned to improve the overall system spectrum efficiency by enabling both information-level and energy-level cooperation \cite{Xu2013CoMP}. Apart from the natural energy sources such as solar and wind that is intermittent due to the environmental change, ambient radio signal has recently been exploited as a new viable source for wireless energy harvesting (WEH) (see \cite{Bi2016overview} and references therein). \textcolor{black}{RF-enabled WET has many preferred advantages. For example, compared with other induction-based WET technologies, WEH can power wireless devices to relatively longer distance exploiting the far-field propogation properties of electromagnetic (EM) wave (e.g., commercial chips available for tens of microwatts ($\mu$W) RF power transferred over $12$m \cite{Powercast}), while the associated transceiver designs are more flexible with the transmitting power, waveforms, occupied resource blocks fully controlled to accommodate different physical conditions.} Joint information and energy cooperation in CR networks has thus been actively investigated in many WEH-enabled scenarios, e.g.,  \cite{Lee13CRN,zheng2014information,Lee15CognitiveWPT,Zhai16TVT}.

The benefit of radio frequency (RF)-powered CCRN is nevertheless compromised by the low wireless power transfer (WPT) efficiency mainly due to the severe RF signal attenuation over distance. One way to improve the WPT efficiency is to employ multiple antennas at the ST, which can improve the EH efficiency of the secondary system \cite{zheng2014information}. The other way to boost the WPT efficiency is to power the ST via WPT from the dedicated energy/hybrid access point (EAP/HAP) \cite{Lee15CognitiveWPT} in addition to the PU.
However, the above works all assume that the involved devices operate in half-duplex (HD) mode, which provides more reliable power supplies for CCRN at the expense of some spectrum efficiency. In continuous effort to address this issue, full-duplex (FD)-enabled communications with wireless information and power transfer has sparked an upsurge of interest thanks to the advance in antenna technologies (see \cite{Ju14FD,Zhong14FD,Zeng15FD,Zheng2013FD,kangEH} and references therein).

In this paper, we consider a spectrum-sharing decode-and-forward (DF) relaying CR network consisting of one pair of primary transmitter (PT) and primary receiver (PR), and one pair of multiple-input multiple-output (MIMO) secondary users (SUs). A number of multi-antenna FD EAPs are coordinated to transfer wireless power to the ST while simultaneously listening to information sent from the PT in the first transmission phase, and decode then forward the PT's message to the PR in HD mode in the second transmission phase. The ST is required to assist the primary transmission and earn the rights to access the PU's spectrum in return; the EAPs are paid by the system as an incentive to support the cooperative WPT and wireless information transfer (WIT). We assume that there is no direct link between the PT and the PR due to severe pathloss \cite{Zheng2013FD}, \textcolor{black}{and the perfect\footnote{We assume perfect CSI at the Tx in this paper as in \cite{zheng2014information,Lee15CognitiveWPT,Ju14FD,Zhong14FD}, and thus the proposed transmission protocol design yields theoretical upper-bound. More practical design for wireless powered MIMO communication taking channel uncertainties into account can be referred to \cite{Boshkovska2017robust} and references therein.} global channel state information (CSI) known at a centralized coordination point who is in charge of acquiring global CSI from the dedicated nodes\footnote{The dedicated nodes are assumed to be those non energy-limited thereby performing channel estimation in line with \cite{Zeng2015training} and then reporting the corresponding CSI to the centralized coordinator connected by backhauls.} via backhauls and implementing the algorithm accordingly in every transmission block (assumed to be equal to the channel coherence time.}  

\textcolor{black}{Compared with \cite{Lee13CRN} investigating joint opportunistic EH and spectrum access, we focus on {\em overlay} CR transmission, which allows for primary messages known at the ST {\em a priori} due to the first-slot transmission, so that the ST can precancel the interference caused to the secondary receiver (SR) by some non-linear precoding techniques, e.g., dirty-paper coding
(DPC). In this case, beamforming precoding for the primary and the secondary messages at the ST needs to be jointly designed to achieve cooperative gain. Furthermore, although
overlay cognitive WPCN has been considered in \cite{Lee15CognitiveWPT} with dedicated WPT, the HAP was only equipped with one antenna therein, and therefore their transmission policy
is not applicable to ours with multi-antennas. In \cite{zheng2014information}, the multi-antenna ST received information from the primary transmitter (PT) and was also fed with energy by the PT using PS and/or time switching (TS) receiver. However, the energy received by the ST was not intended for WPT and thus the RF EH capability and the cooperative gains was limited. By contrast, the deployment of cooperative FD-enabled EAPs intended for WPT in this paper breaks this bottleneck. A wireless powered communication network (WPCN) with an FD-enabled HAP and a set of WEH-enabled time division multiple access (TDMA) users was investigated in \cite{kangEH}. However perfect self interference (SI) cancellation between the transmitting and receiving antennas of the HAP was assumed therein, which is nevertheless not achievable in practice even with the state-of-the-art FD technique \cite{Sabharwal2014JSAC}.} 

A similar setup was considered in \cite{xing2016GC}, whereas our work differs from it mainly in two folds. First, the considered EAPs in this paper are FD empowered so that they fundamentally improve the spectral efficiency of the CCRN system of interest. Second,compared with the non-cooperative EAPs whose power levels are binary (on or off) in \cite{xing2016GC}, we exploit EAP-assisted cooperation in both WPT and WIT phases via continuous power control, which is an extension to the non-cooperative model. The main contributions of this paper are summarized as follows.
\begin{itemize}
\item The weighted sum-rate of the FD EAPs-aided CCRN system is maximized using successive convex approximation (SCA) techniques subject to per-EAP power constraints for WPT and WIT, respectively, the ST's transmitting power constraint, and a practical cost budget that constrains the payment made to the EAPs for their dedicated WPT and WIT.
\item The centralized optimization enables cooperation among the EAPs to effectively mitigate the interference with ST's information decoding (ID), and the SI that degrades EAPs' reception of the PT's signal.
\item A low-complexity suboptimal design locally nulling out the SI at the EAPs is also developed in order to reduce the computational complexity of the iterative algorithm, \textcolor{black}{and is validated by computer simulations to yield performance with little gap to that achieved by the proposed iterative solutions}.
\item Various tradeoffs, e.g., priority between primary and secondary transmissions, energy and cost allocations between WPT and WIT, are studied by solving the optimization problems, and evaluated by simulations to provide useful insights for system design in practice.
\end{itemize}

The remainder of the paper is organized as follows: Section \ref{sec:System Model} and \ref{sec:Problem Formulation} introduces the system model of the CCRN assisted by FD-enabled EAPs, and formulates the weighted sum-rate maximization problem, respectively. Section \ref{sec:Joint Optimization of FD} investigates the feasibility of a tractable reformulation of the original problem; proposes an SCA-based iterative solution along with a suboptimal scheme based upon zero-forcing (ZF) the SI. Benchmark schemes are studied in Section \ref{sec:Benchmark Schemes}. Section \ref{sec:Numerical Results} provides numerical results comparing the performance achieved by different schemes. Finally, Section \ref{sec:Conclusion} concludes the paper.

{\it Notation}---We use the upper case boldface letters for matrices and lower case boldface letters for vectors. $(\cdot)^{T}$, $(\cdot)^{H}$, and ${\sf Tr}(\cdot)$ denote the transpose, conjugate transpose, and trace operations on matrices, respectively. $\|\cdot\|_p$ is \(\ell^p\)-norm of a vector with \(p=2\) by default. The Kronecker product of two matrices is denoted by  $\otimes$. $\mv{A}\succeq 0$ indicates that $\mv{A}$ is a positive semidefinite (PSD) matrix, and $\mv{I}$ denotes an identity matrix with appropriate size. $\mathbb{E}[\cdot]$ stands for the statistical expectation of a random variable (RV). In addition,  $\mathbb{C(R)}^{x \times y}$ stands for the field of complex (real) matrices with dimension $x\times y$, and $\mathbb{Z}$ is the set of integer. $(\cdot)^\ast$ means the optimum solution.

\section{System Model}\label{sec:System Model}
\begin{figure}[htp]
\begin{center}
\includegraphics[width=3.0in]{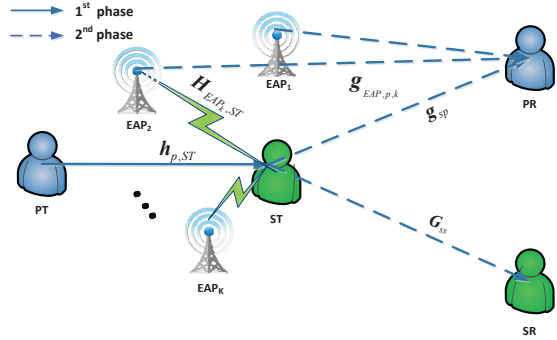}
 \end{center}
 \vspace{-0.3in}
\caption{System model for the wireless powered CCRN.}\label{fig:system model}
\vspace{-0.2in}
\end{figure}

In this paper, we consider a WEH-enabled CCRN that consists of one primary transmitter-receiver pair, one secondary transmitter-receiver pair, and a set of FD-enabled EAPs denoted by \(\mathcal{K}=\{1,\cdots,K\}\) as shown in Fig.~\ref{fig:system model}. The PT and the PR are equipped with one antenna each, while the ST and the SR are equipped with $N$ and $M$ antennas, respectively. The number of transmitting and receiving antennas at the $k$th EAP are denoted by $N_{T,k}$ and $N_{R,k}$, respectively, $\forall k \in \mathcal{K}$, and $N_k=N_{T,k}+N_{R,k}$. We assume that the ST is batteryless, and thus it resorts to WEH as its only source of power for information transmission\textcolor{black}{\footnote{In this paper, we assume that the main energy consumption at the wireless powered ST is from its cooperative transmission, and thereby some other power deletion such as circuit operation \cite{Derrick13OFDMA}, encoding/decoding and pilot transmission are ignored for simplicity of exposition. In addition, we confine our analysis to one transmission block, i.e., channel coherence time, w/o taking the initial energy storage \cite{Wu16EE,Wu16user-centric} into account. However, the solutions developed in the sequel are readily extended to accommodate constant circuit power consumption and/or initial power storage in more practical scenarios.}}.

\begin{figure}[htp]
\begin{center}
\includegraphics[width=3.8in]{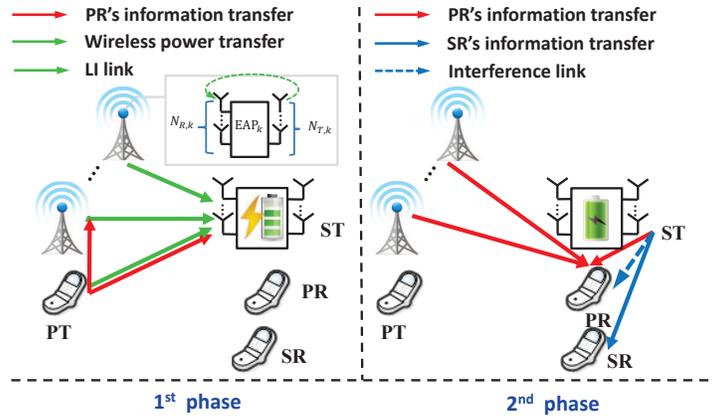}
 \end{center}
\caption{Transmission protocol for the wireless powered CCRN.} \label{fig:transmission protocol}
\vspace{-0.2in}
\end{figure}

As illustrated by Fig.~\ref{fig:transmission protocol}, a two-slot (with equal length) transmission protocol is assumed to be adopted. In the first time slot, the PT transfers the energy-bearing primary user's signal to the ST. Concurrently, the EAPs operating in FD mode cooperate to transfer wireless power to the ST using \(N_{T,k}\)'s antennas, while jointly receiving information from the PT using \(N_{R,k}\)'s antennas. In the secondary time slot, the ST decodes and forwards PT's message and superimposes it on its own to broadcast to the PR and the SR. Meanwhile, the decoded PT's information is also forwarded to the PR by the EAPs that employ \(N_k\) antennas each for information transmission. Let \(s\) denote PT's transmitting signal that follows the circularly symmetric complex Gaussian (CSCG) distribution, denoted by \(s\sim\mathcal{CN}(0,1)\), and \(\mv x\sim\mathcal{CN}(\mv 0, \mv X)\), the energy signals\textcolor{black}{\footnote{Although the CSCG distributed energy signals are not optimal in term of pure WPT, we assume such distribution to exempt the dual-function EAPs from frequent switch between the CSCG and other possible signal generation and to keep its consistence with WIT.}} coordinatedly transmitted by \(K\) EAPs, where \(\mv X\) is the covariance matrix of \(\mv x\). On the other hand, \(\mv x\) can also be alternatively expressed by \(\mv x=[\mv x_k]_{k=1}^K\), where \(\mv x_k\in\mathbb{C}^{N_{T,k}\times 1}\) is the energy signal transmitted by each individual EAP, and is subject to per-EAP power constraint given by \(\mathbb{E}\left[\|\mv x_k\|^2\right]\le P_0\), \(\forall k\in\mathcal{K}\).

\subsection{The First Time Slot}
\textbf{Received signal at the ST.} In this paper, we assume that the ST employs a dynamic power splitting (DPS) receiver \cite{Zhou2013SWIPT} for EH and information decoding (ID) from the same stream of received signal, where \(\varrho\) portion of the received signal power is used to feed the energy supply while the remaining \(1-\varrho\) for ID. As a result, the signal received by the ST is given by
\begin{align}
\mv y_{ST}^{(1)}=\sqrt{1-\varrho}(\mv h_{p,ST}\sqrt{P_p}s+\mv H_{EAP,ST}\mv x+\mv n_a)+\mv n_{c}, \label{eq:first receive at the ST}
\end{align}
where \(\mv h_{p, ST}\in\mathbb{C}^{N\times 1}\) denotes the complex channel from the PT to the ST; \(\mv H_{EAP,ST}=[\mv H_{EAP_1,ST},\cdots,\mv H_{EAP_K,ST}]\); $\mv n_a$ denotes the additive white Gaussian noise (AWGN) at the antennas in RF-band with zero mean and variance $\sigma_{n_a}^2$; $\mv n_c$ is the RF-band to baseband signal conversion noise denoted by \(\mv n_c\sim\mathcal{CN}(\mv 0, \sigma_{n_c}^2\mv I)\). Furthermore, assuming that the linear receiving beamforming performed by the ST is \(\mv u_1^H\mv y_{ST}^{(1)}\), where \(\mv u_1\in\mathbb{C}^{N\times 1}\) is given by maximizing ST's signal-to-interference-plus-noise ratio (SINR) as follows.
\begin{align}
\mv u_1^\ast=\arg\max\limits_{\mv u_1}\frac{(1-\varrho)P_p\vert\mv u_1^H\mv h_{p,ST}\vert^2}{\mv u_1^H((1-\varrho)(\mv H_{EAP,ST}\mv X\mv H_{EAP,ST}^H+\sigma_{n_a}^2\mv I)+\sigma_{n_c}^2\mv I)\mv u_1}, \label{eq:first-slot SINR at the ST}
\end{align}
which proves to be the eigenvector corresponding to the largest (generalized) eigenvalue of matrices \(((1-\varrho)(\mv H_{EAP,ST}\mv X\mv H_{EAP,ST}^H+\sigma_{n_a}^2\mv I)+\sigma_{n_c}^2\mv I,\mv h_{p,ST}\mv h_{p,ST}^H)\). It thus leads to the SINR at the  ST in the first transmission phase given by
\begin{align}
\lambda_{\max}((1-\varrho)P_p\mv A^{-\frac{1}{2}}\mv h_{p,ST}\mv h_{p,ST}^H\mv A^{-\frac{1}{2}})\stackrel{\rm(a)}{=}&\lambda_{\max}((1-\varrho)P_p\mv h_{p,ST}^H\mv A^{-1}\mv h_{p,ST})\notag\\
=&(1-\varrho)P_p\mv h_{p,ST}^H\mv A^{-1}\mv h_{p,ST}, \label{eq:compact first-slot SINR at the ST}
\end{align} where \(\mv A=(1-\varrho)(\mv H_{EAP,ST}\mv X\mv H_{EAP,ST}^H+\sigma_{n_a}^2\mv I)+\sigma_{n_c}^2\mv I\), and \(\rm (a)\) is due to the rank-one \(\mv A^{-\frac{1}{2}}\mv h_{p,ST}\mv h_{p,ST}^H\mv A^{-\frac{1}{2}}\).

\textbf{Received signal at the EAPs.} The received signal at the EAPs that is interfered with the energy signals transmitted by the same EAPs can be expressed in a vector form given by
\begin{align}
\mv y_{EAP}=\mv h_{p,EAP}\sqrt{P_p}s+\textcolor{black}{{\mv H}_{TR}\mv x}+\mv n_{EAP}^{(1)}, \label{eq:first receive at the EAPs}
\end{align} where \(\mv h_{p,EAP}=[\mv h_{p,EAP_1}^H,\cdots,\mv h_{p,EAP_K}^H]^H\) with \(\mv h_{p,EAP_k}\in\mathbb{C}^{N_{R,k}\times 1}\), \(k\in\mathcal{K}\),  denoting the complex channel from the PT to the \(k\)th EAP; \({\mv H}_{TR}\) indicates the effective loop interference (LI) channel from the transmitting to the the receiving antennas of the EAPs after analogue domain SIC, in which  the block matrices \(\mv H_{T_k,R_k}\in \mathbb{C}^{N_{R,k}\times N_{T,k}}\) on the diagonal of \(\mv H_{TR}\), \(\forall k\), denote the intra-EAP LI  channels from within the \(k\)th EAP, and the matrices \(\mv H_{T_i,R_j}\) off the diagonal, \(\forall i\neq j\), represent the inter-EAP LI channels from the \(i\)th EAP to the \(j\)th EAP; \(\mv n_{EAP}^{(1)}\) is assumed to be the AWGN noise received at the EAPs, i.e., \(\mv n_{EAP}^{(1)}\sim\mathcal{CN}(\mv 0,\sigma_{EAP}^2\mv I)\).
Without loss of generality (w.l.o.g.), \(\mv H_{TR}\) is given by \(\mv H_{TR}=\sqrt{\varphi^2}\bar{\mv H}_{TR}\), where each element in \(\mv H_{T_k,R_k}\)'s is a complex Gaussian RV with zero mean and variance of \(\varphi^2\); \textcolor{black}{each element in \(\mv H_{T_i,R_j}\)'s is a complex Gaussian RV with zero mean and variance of \(\varphi^2\) multiplied by path-loss}. \(\bar{\mv H}_{TR}\) denotes the normalized complex LI channel, and \(\varphi^2\in[0,1]\) indicates the residual LI channel gains.

It is worth noting that the analogue domain SIC is implemented though, the power level of the residual LI can still be much larger than that of the desired signal \cite{Kim13FD}, i.e., \(\varphi^2\mathbb{E}[\|\bar{\mv H}_{TR}\mv x\|^2]\gg P_p\|\mv h_{p,EAP}\|^2\), due to which the following concern arises. Channel estimation errors w.r.t the block diagonal matrices \(\bar{\mv H}_{T_k,R_k}\)'s cannot be neglected. Assume that the estimation of \(\bar{\mv H}_{TR}\) is given by \(\bar{\mv H}_{TR}=\hat{\mv H}_{TR}+\sqrt{\varepsilon^2}\tilde{\mv H}_{TR}\) \cite{Ju14FD}, where \(\hat{\mv H}_{TR}\) denotes the estimation of  \(\bar{\mv H}_{TR}\); \(\tilde{\mv H}_{TR}\) denotes its errorenous channel, whose elements are complex Gaussian RVs with zero mean and variance of \(1\) in the block matrices \(\tilde{\mv H}_{T_k,R_k}\)'s, \textcolor{black}{and variance of path-loss in the block matrices \(\mv H_{T_i,R_j}\)'s}, respectively. \(\epsilon^2\ll 1\) denotes the level of estimation accuracy. Hence, after analogue-to-digital conversion (ADC) \cite{Day12FD}, digital domain SIC is further applied to subtract \(\sqrt{\varphi^2}\hat{\mv H}_{TR}\) from \eqref{eq:first receive at the EAPs}. The processed signal is accordingly expressed as
\begin{align}
\bar{\mv y}_{EAP}=\mv h_{p,EAP}\sqrt{P_p}s+\sqrt{\varphi^2\varepsilon^2}\tilde{\mv H}_{TR}\mv x+\mv n_{EAP}^{(1)}. \label{eq:processed receive at the EAPs}
\end{align}
For the purpose of exposition, we denote \(\varphi\varepsilon\) by \(\theta\) in the sequel.

Furthermore, since the \(K\) EAPs are coordinated, they can perform joint decoding of the PT's signal \(s\) to maximize their receiving SINR. Therefore the optimum receiving beamforming \(\mv u_2\) is designed such that
\begin{align}
SINR_{EAP}&=\max\limits_{\mv u_2}\frac{P_p\vert\mv u_2^H\mv h_{p,EAP}\vert^2}{\mv u_2^H(\theta^2\tilde{\mv H}_{TR}\mv X\tilde{\mv H}_{TR}^H+\sigma_{EAP}^2\mv I)\mv u_2},\label{eq:first-slot SINR at the EAPs}
\end{align}
which is equal to \(\lambda_{\max}(P_p\mv B^{-\frac{1}{2}}\mv h_{p,EAP}\mv h_{p,EAP}^H\mv B^{-\frac{1}{2}})\), where \(\mv B=\theta^2\tilde{\mv H}_{TR}\mv X\tilde{\mv H}_{TR}^H+\sigma_{EAP}^2\mv I\).

\subsection{The Second Time Slot}
\textbf{Transmitted signal at the ST.} In the second time slot, the ST extracts the PR's desired message and superimposes it with its own message using dirty-paper coding (DPC) as follows \cite{Caire2003throughput}.
\begin{align}
  \mv x_{ST}^{(2)}=\mv w_ps+\mv q_s, \label{eq:transmit at the ST}
\end{align}
where \(\mv w_p\) is the beamforming vector for \(s\), while \(\mv q_s\) is the transmitted signal conveying the SR's information aimed for multiplexing  MIMO transmission\textcolor{black}{\footnote{In fact, \(\mv w_ps\) and \(\mv q_s\) are transmitting signals precoded by DPC first and then multi-antenna beamforming.}}, the covariance matrix of which is \(\mathbb{E}[\mv q_s\mv q_s^H]=\mv Q_s\). As mentioned before, the transmitting power for the ST is solely supplied by its harvest power, i.e.,
\begin{align}
{\sf Tr}(\mv Q_s)+\|\mv w_p\|^2\mv\le\eta\varrho P_{\rm EH}(\mv X), \label{eq:harvested power constraint}
\end{align}
where \(P_{\rm EH}(\mv X)={\sf Tr}(\mv H_{EAP,ST}\mv X\mv H_{EAP,ST}^H)+P_p\|\mv h_{p,ST}\|^2\) is the total wireless transferred power, and $\eta$ denotes the EH conversion  efficiency\textcolor{black}{\footnote{In this paper, we focus on the transceiver design of the wireless powered ST and the FD-enabled EAPs assuming a simple linear EH model. The interested reader can refer to \cite{Boshkovska15nonlinear,Dong16nonlinear} for non-linear EH modelling taking the dependence of \(\eta\) on the input harvested power into account.}}.

\textbf{Transmitted signal at the EAPs.} In the second time slot, the EAPs cooperatively transmit the decoded PT's message to the PR using all of their \(N_k\)'s antennas  via beamforming \(\mv v_ps\), where \(\mv v_p=[\mv v_{p,1}^H, \cdots, \mv v_{p,K}^H]^H\), and \(v_{p,k}\in\mathbb{C}^{N_k\times 1}\), \(k\in\mathcal{K}\),  represents the \(k\)the EAP's  beamforming vector.

\textbf{Received signal at the PR.} In the second time slot, PR receives the forwarded PT's message from both the ST and EAPs as follows.
\begin{align}
  y_{PR}^{(2)}=\mv g_{sp}^H\mv x_{ST}^{(2)}+\mv g_{EAP,p}^H\mv v_ps+n_{PR}^{(2)},  \label{eq:second receive at the PR}
\end{align}
where \(\mv g_{sp}\in\mathbb{C}^{N\times 1}\) is the Hermitian transpose of the complex channels from the ST to the PR, \(\mv g_{EAP,p}=[\mv g_{EAP_1,p}^H, \cdots,\mv g_{EAP_K,p}^H ]^H\) with \(g_{EAP_k,p}\in\mathbb{C}^{N_k\times 1}\), \(k\in\mathcal{K}\),  is the Hermitian transpose of those from the EAPs to the PR, and $n_{PR}^{(2)}$ is the AWGN at the PR denoted by \(n_{PR}^{(2)}\sim\mathcal{CN}(0, \sigma_{PR}^2)\). Plugging \eqref{eq:transmit at the ST} into \eqref{eq:second receive at the PR}, \(y_{PR}^{(2)}\) can be rewritten as
\begin{align}
y_{PR}^{(2)}=(\mv g_{sp}^H\mv w_p+\mv g_{EAP,p}^H\mv v_p)s+\mv g_{sp}^H\mv q_s+n_{PR}^{(2)}.  \label{eq:compact y_PR(2)}
\end{align}
The receiving SINR for the PR treating the interference caused by the secondary information as noise in the second transmission slot is thus given by
\begin{align}
{\rm SINR}_{PR}=\frac{\left\vert\mv g_{sp}^H\mv w_p+\mv g_{EAP,p}^H\mv v_p\right\vert^2}{\mv g_{sp}^H\mv Q_s\mv g_{sp}+\sigma_{PR}^2}. \label{eq:SINR at the PR}
\end{align}
Accordingly, the achievable DF relaying rate for the PR, denoted by \(r_{PR}(\mv X,\varrho)\), is given by
\begin{multline}
r_{PR}(\mv X,\varrho)=\min\Bigg\{\max\Big\{\frac{1}{2}\log_2\left(1+(1-\varrho)P_p\mv h_{p,ST}^H\mv A^{-1}(\varrho,\mv X)\mv h_{p,ST}\right),\\
\frac{1}{2}\log_2\left(1+P_p\mv h_{p,EAP}^H\mv B^{-1}(\mv X)\mv h_{p,EAP}\right)\Big\},
\frac{1}{2}\log_2\left(1+\frac{\left\vert\mv g_{sp}^H\mv w_p+\mv g_{EAP,p}^H\mv v_p\right\vert^2}{\mv g_{sp}^H\mv Q_s\mv g_{sp}+\sigma_{PR}^2}\right)\Bigg\}. \label{eq:achievable rate for the PR}
\end{multline}

\textbf{Received signal at the SR.} \textcolor{black}{Assuming that DPC is adopted at the ST, by which the ST encodes its own message with the interference caused by the PT's message known {\em a priori}, the SR is able to receive no interference as follows.
\begin{align}
\mv y_{SR}^{(2)}=\mv G_{ss}\mv q_s+\mv n_{SR}^{(2)}, \label{eq:second receive at the SR}
\end{align}
where \(\mv G_{ss}\) denotes the MIMO channels between the ST and the SR and \(\mv n_{SR}^{(2)}\) is the received noise at the SR, denoted by \(\mv n_{SR}^{(2)}\sim\mathcal{CN}(0,\sigma_{SR}^2\mv I)\). Accordingly, the achievable rate for the secondary overlay MIMO transmission, denoted by \(r_{SR}\), is given by
\begin{align}
r_{SR}(\mv Q_s)=\frac{1}{2}\log_2\det\left(\mv I+\frac{\mv G_{ss}\mv Q_s\mv G_{ss}^H}{\sigma_{SR}^2}\right). \label{eq:achievable rate for the SR}
\end{align}}

\section{Problem Formulation}\label{sec:Problem Formulation}
\textcolor{black}{In this paper, we assume that the spectrum sharing CCRN of interest aims for maximizing the weighted sum-rate, i.e., \(c_1r_{PR}(\mv X,\varrho)+c_2r_{SR}(\mv Q_s)\) (c.f.~\eqref{eq:achievable rate for the PR} and \eqref{eq:achievable rate for the SR}), where \(c_1\) and \(c_2\) are weight coefficients that balance the priority of service between the primary and secondary system.} \textcolor{black}{Since the ST is required to assist with the primary transmission by DF relaying using its harvested power from the EAPs and the PT, we assume that the EAPs charge \(c_3 \eta {\sf Tr}(\mv H_{EAP,ST}\mv X\mv H_{EAP,ST}^H)\) from the ST for its harvested power, where \(c_3\) is a cost conversion factor. Moreover, the EAPs also collaborate to help relaying PT's message thus alleviating the burden of the energy-limited ST. As a return, the ST pays the EAPs an amount of \(c_4\vert \mv g_{EAP,p}^H\mv v_p\vert^2\) for their information transmission, where \(c_4\) represents the cost per unit of received PT's signal power (c.f.~\eqref{eq:second receive at the PR}).}
In summary, the total cost for the FD-enabled EAPs-aided CCRN is constrained by \(c_3\eta\varrho{\sf Tr}(\mv H_{EAP,ST}\mv X\mv H_{EAP,ST}^H) +c_4\vert \mv g_{EAP,p}^H\mv v_p\vert^2\le C\), where \(C\) is the total budget of the ST. 

\textcolor{black}{It is worth noting that this constraint will have a impact on the system only when \(C\le C_{\max}\), where \(C_{\max}\) denotes the maximum possible system payment given by the following problem.
\begin{align*}
\mathrm{(P0)}:\mathop{\mathtt{max}}_{\mv X, \mv v_p}
&~~~c_3\eta\varrho{\sf Tr}(\mv H_{EAP,ST}\mv X\mv H_{EAP,ST}^H)+c_4\vert\mv g_{EAP,p}^H\mv v_p\vert^2\\
\mathtt {s.t.}&~~~{\sf Tr}(\mv E_k\mv X)\le P_0, \; \forall k,\\
&~~~\|\mv v_{p,k}\|^2\le P_0, \; \forall k.
\end{align*}}
Note that the addends of the objective function of Problem \(\mathrm{(P0)}\) are independent of each other and thus can be solved separately. Specifically, the second term \(\vert\mv g_{EAP,p}^H\mv v_p\vert^2\), which accounts for the amount of PT's signal power received by the PR, yields a closed-form solution shown below. 
\begin{align}
\vert \mv g_{EAP,p}^H\mv v_p\vert^2
\stackrel{\rm (a)}{\le} &\left(\sum_{k=1}^K\left\vert\mv g_{EAP_k,p}^H\mv v_{p,k}\right\vert\right)^2
\stackrel{\rm (b)}{\le} &\left(\sqrt{P_0}\sum_{k=1}^K\left\| \mv g_{EAP_k,p} \right\|_2\right)^2
=P_0\left\|\tilde{\mv g}_{EAP,p}\right\|_1^2, \label{eq:tilde g_{EAP,p}}
\end{align} where the equality in \({\rm (a)}\) holds when all \(\mv g_{EAP,p,k}^H\mv v_{p,k}\)'s are aligned in the same direction; \({\rm (b)}\) is due to Cauchy-Schwartz inequality; and \(\tilde{\mv g}_{EAP,p}=\left[\|\mv g_{EAP_1,p}\|_2, \cdots, \|\mv g_{EAP_K,p}\|_2\right]^T\). As a result, only \(\mv X\succeq\mv 0\) remains to be solved by \({\kern-4pt}\mathop{\mathtt{max}}_{{\sf Tr}(\mv E_k\mv X)\le P_0,\; \forall k}{\kern-4pt}{\sf Tr}(\mv H_{EAP,ST}\mv X\mv H_{EAP,ST}^H)\), the optimum solution of which is denoted by \(\mv X^\ast\). Thus, \(C_{\max}\) turns out to be \(c_3\eta\varrho{\sf Tr}(\mv H_{EAP,ST}\mv X^\ast\mv H_{EAP,ST}^H)+c_4P_0\|\tilde{\mv g}_{EAP,p}\|_1^2\).

Next, the weighted sum-rate maximization problem subject to the harvested power at the ST, the total payment charged by the EAPs, and the per-EAP power constraints can be formulated as follows.
\begin{subequations}
\begin{align}
\mathrm{(P1)}: \kern-4pt\mathop{\mathtt{max}}_{\mv X, \mv Q_s, \mv w_p, \mv v_p,  \varrho}\kern-4pt
&~~~c_1r_{PR}(\mv X,\varrho)+c_2r_{SR}(\mv Q_s)\notag\\
\mathtt {s.t.}& ~~~\|\mv v_{p,k}\|^2\le P_0, \; \forall k, \label{eq:per-EAP power constraint for WIT}\\
&~~~{\sf Tr}(\mv E_k\mv X)\le P_0, \; \forall k, \label{eq:per-EAP power constraint for WPT in terms of X}\\
&~~~{\sf Tr}(\mv Q_s)+\|\mv w_p\|^2\le\eta\varrho P_{\rm EH}(\mv X), \label{eq:sum power constraint}\\
&~~~c_3\eta\varrho{\sf Tr}(\mv H_{EAP,ST}\mv X\mv H_{EAP,ST}^H)+c_4\vert \mv g_{EAP,p}^H\mv v_p\vert^2\le C, \label{eq:sum cost constraint}\\
&~~~0\le\varrho\le 1, \label{eq:constraint for varrho}\\
&~~~\mv X\succeq\mv 0, \, \mv Q_s\succeq\mv 0,
\end{align}
\end{subequations} where \(\mv E_k={\rm Diag}([\mv 0, \cdots, \mv I_k, \cdots, \mv 0])\), in which \({\rm Diag}([\cdot])\) denotes a block diagonal matrix with the block matrices on the diagonal given in \([]\).

In the above problem \(\mathrm{(P1)}\), \eqref{eq:per-EAP power constraint for WIT} and \eqref{eq:per-EAP power constraint for WPT in terms of X} illustrates the per-EAP power constraint for information and power transfer, respectively; \eqref{eq:sum power constraint} indicates the transmitting power constraint of the ST subject to its harvested power from the EAPs and the PT; and \eqref{eq:sum cost constraint} constrains the cost of the CCRN system no more than a constant value \(C\).

\section{Joint Optimization of FD Energy Beamforming and DF Relaying}\label{sec:Joint Optimization of FD}
\textcolor{black}{In this section we investigate to solve problem \(\mathrm{(P1)}\). First, we remove the inner \(\max(\cdot)\) in \(r_{PR}(\mv X,\varrho)\) (c.f.~\eqref{eq:achievable rate for the PR}) by recasting \(\mathrm{(P1)}\) into two subproblems, denoting the optimum value of  \(\mathrm{(P1)}\) by \((f^\ast=\max{\left\{f_1^\ast,f_2^\ast\right\}}\), where \(f_1^\ast\) and \(f_2^\ast\) are the optimum value of subproblems \(\mathrm{(P1.1)}\) and \(\mathrm{(P1.2)}\), respectively.} They represent the cases in which the transmission rate of the first-slot DF relaying is achieved by the ST and the EAPs, respectively, depending on whether \((1-\varrho)\mv h_{p,ST}^H\mv A^{-1}(\varrho,\mv X)\mv h_{p,ST}\) is larger than \(\mv h_{p,EAP}^H\mv B^{-1}(\mv X)\mv h_{p,EAP}\) or not.

Problem \(\mathrm{(P1.1)}\) based on the epigraph reformulation is given by
\begin{subequations}
\begin{align}
\mathrm{(P1.1)}: \kern-4pt\mathop{\mathtt{max}}_{\mv X, \mv Q_s, \mv w_p, \mv v_p, \varrho, t}\kern-4pt
&~~~\frac{1}{2}c_1\log_2(1+t)+c_2r_{SR}(\mv Q_s)\notag\\
\mathtt {s.t.}& ~~~\eqref{eq:per-EAP power constraint for WIT}\--\eqref{eq:constraint for varrho},\\
&~~~(1-\varrho)P_p\mv h_{p,ST}^H\mv A^{-1}(\varrho,\mv X)\mv h_{p,ST}\ge P_p\mv h_{p,EAP}^H\mv B^{-1}(\mv X) h_{p,EAP}, \label{eq:SINR of the ST superior to that of the EAPs}\\
&~~~(1-\varrho)P_p\mv h_{p,ST}^H\mv A^{-1}(\varrho,\mv X)\mv h_{p,ST}\ge t, \label{eq:SINR of the ST larger than t}\\
&~~~\frac{\left\vert\mv g_{sp}^H\mv w_p+\mv g_{EAP,p}^H\mv v_p\right\vert^2}{\mv g_{sp}^H\mv Q_s\mv g_{sp}+\sigma_{PR}^2}\ge t, \label{eq:minimum PR's SINR threshold}\\
&~~~\mv X\succeq\mv 0, \, \mv Q_s\succeq\mv 0, \, t\ge 0, \label{eq:constraint for PSD variables and t}
\end{align}
\end{subequations} where \(\mv A\) is related to the optimization variables \(\varrho\) and \(\mv X\), and thus denoted by \(\mv A(\varrho,\mv X)\) for the convenience of exposition, while \(\mv B\) is denoted by \(\mv B(\mv X)\).

\(\mathrm{(P1.2)}\) is similarly given by
\begin{subequations}
\begin{align}
\mathrm{(P1.2)}: \kern-4pt\mathop{\mathtt{max}}_{\mv X, \mv Q_s, \mv w_p, \mv v_p, \varrho, t}\kern-4pt
&~~~\frac{1}{2}c_1\log_2(1+t)+c_2r_{SR}(\mv Q_s)\notag\\
\mathtt {s.t.}& ~~~\eqref{eq:per-EAP power constraint for WIT}\--\eqref{eq:constraint for varrho},\ \eqref{eq:minimum PR's SINR threshold}\--\eqref{eq:constraint for PSD variables and t},\\
&~~~P_p\mv h_{p,EAP}^H\mv B^{-1}(\mv X)\mv h_{p,EAP}\ge (1-\varrho)P_p\mv h_{p,ST}^H\mv A^{-1}(\varrho, \mv X)\mv h_{p,ST}, \label{eq:SINR of the EAPs superior to that of the ST}\\
&~~~P_p\mv h_{p,EAP}^H\mv B^{-1}(\mv X)\mv h_{p,EAP}\ge t.\label{eq:SINR of the EAPs larger than t}
\end{align}
\end{subequations}

\subsection{Problem Reformulation}
\textcolor{black}{It is observed that \(\mv X\) and \(\varrho\) are coupled together in \eqref{eq:SINR of the ST superior to that of the EAPs}, \eqref{eq:SINR of the ST larger than t}, and \eqref{eq:sum power constraint}, which make these constraints non-convex. Hence, in this subsection, we reformulate the non-convex subproblems in a tractable way.} Specifically, we propose to solve problem \(\mathrm{(P1.1)}\) in two stages as follows. First, given \(\bar\varrho\in[0,1]\), we solve the following problem.
\begin{subequations}
\begin{align}
\mathrm{(P1.1\text{-}1)}: \kern-4pt\mathop{\mathtt{max}}_{\mv X, \mv Q_s, \mv w_p, \mv v_p, t}\kern-4pt
&~~~\frac{1}{2}c_1\log_2(1+t)+c_2r_{SR}(\mv Q_s)\notag\\
\mathtt {s.t.}& ~~~\eqref{eq:per-EAP power constraint for WIT}\--\eqref{eq:per-EAP power constraint for WPT in terms of X},\ \eqref{eq:minimum PR's SINR threshold}\--\eqref{eq:constraint for PSD variables and t},\\
&~~~(1-\bar\varrho)P_p\mv h_{p,ST}^H\mv A^{-1}(\bar\varrho,\mv X)\mv h_{p,ST}\ge P_p\mv h_{p,EAP}^H\mv B^{-1}(\mv X) h_{p,EAP}, \label{eq:SINR of the ST superior to that of the EAPs with fixed rho}\\
&~~~(1-\bar\varrho)P_p\mv h_{p,ST}^H\mv A^{-1}(\bar\varrho,\mv X)\mv h_{p,ST}\ge t,\label{eq:SINR of the ST larger than t with fixed rho}\\
&~~~{\sf Tr}(\mv Q_s)+\|\mv w_p\|^2\mv\le\eta\bar\varrho P_{\rm EH}(\mv X), \label{eq:sum power constraint with fixed rho}\\
&~~~c_3\eta\bar\varrho{\sf Tr}(\mv H_{EAP,ST}\mv X\mv H_{EAP,ST}^H)+c_4\vert \mv g_{EAP,p}^H\mv v_p\vert^2\le C.\label{eq:sum cost constraint with fixed rho}
\end{align}
\end{subequations}
Then, denoting the optimum value of  problem \(\mathrm{(P1.1\text{-}1)}\) as \(f_1(\varrho)\), the optimum \(\varrho\) can be found by solving \(\mathrm{(P1.1\text{-}2)}:\) \(\max\limits_{\varrho\in[0,1]}f_1(\varrho)\) via one-dimension search over \(\varrho\), which guarantees an \(\epsilon\)-optimum\footnote{\(\epsilon\)-optimum means that \(\forall\epsilon>0\), to achieve an objective value in the \(\epsilon\)-neighbourhood of its optimum value, there always exits a corresponding one-dimension search step length.} solution. Hence, we focus on solving \(\mathrm{(P1.1\text{-}1)}\) in the sequel.

Note that since both \(\mv h_{p,ST}^H\mv A^{-1}(\bar\varrho,\mv X)\mv h_{p,ST}\),  denoted by \(g_1(\mv X)\), and \(\mv h_{p,EAP}^H\mv B^{-1}(\mv X) h_{p,EAP}\), denoted by \(g_2(\mv X)\), are proved to be convex functions w.r.t \(\mv X\) (see Appendix~\ref{appendix:proof of convexity of the constraint of ST's receiving SINR superior to that of the EAPs}), the constraints given by \eqref{eq:SINR of the ST superior to that of the EAPs with fixed rho} and \eqref{eq:SINR of the ST larger than t with fixed rho} are in general not convex. However, \eqref{eq:SINR of the ST superior to that of the EAPs with fixed rho} is seen to admit the form of difference of convex (D.C.) functions, which falls into the category of D.C. programming \cite{Yuille03DC}, and thus is solvable by employing D.C. iterations \cite{Kha12DC}. Specifically, we replace the left-hand-side (LHS) of \eqref{eq:SINR of the ST superior to that of the EAPs with fixed rho} (\eqref{eq:SINR of the ST larger than t with fixed rho}) with its first-order Taylor expansion w.r.t. \(\bar{\mv X}\), since it is a global lower-bound estimator of the convex function \(g_1(\mv X)\) and affine. Therefore, \eqref{eq:SINR of the ST superior to that of the EAPs with fixed rho} can be transformed into the following convex constraint.
\begin{align}
(1-\bar\varrho)P_p\left(g_1(\bar{\mv X})+\Re\left\{{\sf Tr}\left(\nabla_{\mv X}g_1(\bar{\mv X})(\mv X-\bar{\mv X})\right)\right\}\right)\ge P_p\mv h_{p,EAP}^H\mv B^{-1}(\mv X) h_{p,EAP}, \label{eq:lower-bound SINR of the ST superior to that of the EAPs with fixed rho}
\end{align}
where \(\nabla_{\mv X}g_1(\mv X)\), given by
\begin{align}
\nabla_{\mv X}g_1(\mv X)=-(1-\bar\varrho)\mv H_{EAP,ST}^H{\mv A}^{-1}(\bar\varrho, \mv X)\mv h_{p,ST}\mv h_{p,ST}^H{\mv A}^{-1}(\bar\varrho, \mv X)\mv H_{EAP,ST},
\end{align} denotes the gradient matrix of \(g_1(\mv X)\). Accordingly, plugging the LHS of \eqref{eq:lower-bound SINR of the ST superior to that of the EAPs with fixed rho} into \eqref{eq:SINR of the ST larger than t with fixed rho}, the  constraint \eqref{eq:SINR of the ST larger than t with fixed rho} is also made convex as follows.
\begin{align}
(1-\bar\varrho)P_p\left(g_1(\bar{\mv X})+\Re\left\{{\sf Tr}\left(\nabla_{\mv X}g_1(\bar{\mv X})(\mv X-\bar{\mv X})\right)\right\}\right)\ge t. \label{eq:lower-bound SINR of the ST larger than t with fixed rho}
\end{align}
It is easy to verify that satisfying constraints \eqref{eq:lower-bound SINR of the ST superior to that of the EAPs with fixed rho} and \eqref{eq:lower-bound SINR of the ST larger than t with fixed rho} implies feasibility of \eqref{eq:SINR of the ST superior to that of the EAPs with fixed rho} and \eqref{eq:SINR of the ST larger than t with fixed rho}, but the converse is not necessarily true. Hence, \eqref{eq:lower-bound SINR of the ST superior to that of the EAPs with fixed rho} and \eqref{eq:lower-bound SINR of the ST larger than t with fixed rho} in general shrink the feasible region of \(\mathrm{(P1.1\text{-}1)}\) unless \(\mv X^\ast=\bar{\mv  X}\), and only lead to its lower-bound solution, which will be discussed in detail later.

Next, we look into the constraint \eqref{eq:minimum PR's SINR threshold}, which is non-convex due to coupling \(\mv Q_s\) and \(t\). To facilitate solving \(\mathrm{(P1.1\text{-}1)}\), we decouple the numerator and the denominator of its LHS by introducing an auxiliary variable \(y>0\) as follows.
\begin{align}
&\vert\mv g_{sp}^H\mv w_p+\mv g_{EAP,p}^H\mv v_p\vert\ge \sqrt{ty}, \label{eq:convex larger than concave}\\
&\mv g_{sp}^H\mv Q_s\mv g_{sp}+\sigma_{PR}^2\le y, \label{eq:maximum permissive interference caused by ST}
\end{align}
which prove to be sufficient and necessary to replace \eqref{eq:minimum PR's SINR threshold}. However, as \(\sqrt{ty}\) is jointly concave w.r.t. \(t\) and \(y\) over \(t>0\) and \(y>0\), \eqref{eq:convex larger than concave} is still non-convex. To accommodate this constraint to the framework of convex optimization, we equivalently transform the LHS of \eqref{eq:convex larger than concave} into a linear form based upon the following lemma.
\begin{lemma}
The optimum value of \(\mathrm{(P1.1)}\) is attained when the solution of \(\mv w_p^\ast\), and \(\mv v_p^\ast\) satisfies \(\measuredangle(\mv g_{sp}^H\mv w_p^\ast)=\measuredangle(\mv g_{EAP,p}^H\mv v_p^\ast)\). \label{lemma:the same direction of two inner-products}
\end{lemma}
\begin{IEEEproof}
Please refer to Appendix~\ref{appendix:proof of the same direction of two inner-produts}.
\end{IEEEproof}

In accordance with Lemma~\ref{lemma:the same direction of two inner-products}, we have the constraint \eqref{eq:convex larger than concave} equivalently expressed as
\begin{align}
\Re\left\{\mv g_{sp}^H\mv w_p+\mv g_{EAP,p}^H\mv v_p\right\}\ge\sqrt{ty}. \label{eq:absolute value removing}
\end{align}
By rotating any solution \(\mv w_p^\ast\) and \(\mv v_p^\ast\) with a common angle of \(-\measuredangle(\mv g_{sp}^H\mv w_p^\ast)\), \eqref{eq:convex larger than concave} turns out to be \eqref{eq:absolute value removing} without violating any other constraints.

To deal with the RHS of \eqref{eq:absolute value removing}, i.e., \(\sqrt{ty}\), since it is jointly w.r.t \(t>0\) and \(y>0\), its first-order Taylor expansion given by
\begin{align}
\sqrt{ty}\le\sqrt{\bar t\bar y}+\frac{1}{2}\sqrt{\frac{\bar y}{\bar t}}(t-\bar t)+\frac{1}{2}\sqrt{\frac{\bar t}{\bar y}}(y-\bar y) \label{eq:Talyor expansion of t and y}
\end{align} serves as its upper-bound approximation, in which the equality holds if and only if \(t=\bar t\) and \(y=\bar y\). Hence, \eqref{eq:convex larger than concave} can be approximated by a convex constraint expressed as
\begin{align}
\Re\left\{\mv g_{sp}^H\mv w_p+\mv g_{EAP,p}^H\mv v_p\right\}\ge \sqrt{\bar t\bar y}+\frac{1}{2}\sqrt{\frac{\bar y}{\bar t}}(t-\bar t)+\frac{1}{2}\sqrt{\frac{\bar t}{\bar y}}(y-\bar y). \label{eq:affine larger than affine}
\end{align}

Finally, the non-convex problem \(\mathrm{(P1.1\text{-}1)}\) is reformulated as the following convex problem.
\begin{subequations}
\begin{align}
\mathrm{(P1.1\text{-}1^\prime)}: \kern-4pt\mathop{\mathtt{max}}_{\mv X, \mv Q_s, \mv w_p, \mv v_p,t,y}\kern-4pt
&~~~\frac{1}{2}c_1\log_2(1+t)+c_2r_{SR}(\mv Q_s)\notag\\
\mathtt {s.t.}& ~~~\eqref{eq:per-EAP power constraint for WIT}\--\eqref{eq:per-EAP power constraint for WPT in terms of X},\ \eqref{eq:constraint for PSD variables and t},\ \eqref{eq:lower-bound SINR of the ST superior to that of the EAPs with fixed rho},\ \eqref{eq:lower-bound SINR of the ST larger than t with fixed rho},\  \eqref{eq:maximum permissive interference caused by ST},\ \eqref{eq:affine larger than affine},\ \eqref{eq:sum power constraint with fixed rho}\--\eqref{eq:sum cost constraint with fixed rho},\\
&~~~y\ge 0. \label{eq:y larger than zero}
\end{align}
\end{subequations}

Problem \(\mathrm{(P1.2)}\) can also be similarly treated and transformed into a two-stage problem, for which we employ the first-order Taylor expansion of the convex function \(g_2(\mv X)\) in the LHS of \eqref{eq:SINR of the EAPs superior to that of the ST} and \eqref{eq:SINR of the EAPs larger than t}, respectively, to serve as their lower-bound approximation. As is done with \eqref{eq:SINR of the ST superior to that of the EAPs with fixed rho} and \eqref{eq:SINR of the ST larger than t with fixed rho}, given any \(\varrho=\bar\varrho\), \eqref{eq:SINR of the EAPs superior to that of the ST} and \eqref{eq:SINR of the EAPs larger than t} can be approximated by
\begin{align}
P_p\left(g_2(\bar{\mv X})+\Re\left\{{\sf Tr}\left(\nabla_{\mv X}g_2(\bar{\mv X})(\mv X-\bar{\mv X})\right)\right\}\right)\ge (1-\bar\varrho)P_p\mv h_{p,ST}^H\mv A^{-1}(\bar\varrho, \mv X) h_{p,ST}, \label{eq:lower-bound SINR of the EAPs superior to that of the ST with fixed rho}
\end{align} and
\begin{align}
P_p\left(g_2(\bar{\mv X})+\Re\left\{{\sf Tr}\left(\nabla_{\mv X}g_2(\bar{\mv X})(\mv X-\bar{\mv X})\right)\right\}\right)\ge t, \label{eq:lower-bound SINR of the EAPs larger than t with fixed rho}
\end{align} respectively, where \(\nabla_{\mv X}g_2(\mv X)\) is given by
\begin{align}
\nabla_{\mv X}g_2(\mv X)=-\theta^2\tilde{\mv H}_{TR}^H{\mv B}^{-1}(\mv X)\mv h_{p,EAP}\mv h_{p,EAP}^H{\mv B}^{-1}(\mv X)\tilde{\mv H}_{TR}.
\end{align} In addition, \(\mathrm{(P1.2)}\) shares the same constraint \eqref{eq:minimum PR's SINR threshold} with \(\mathrm{(P1.1)}\), which can be approximated by the same constraints \eqref{eq:maximum permissive interference caused by ST} and \eqref{eq:affine larger than affine}. Hence, the corresponding main stage of solving \(\mathrm{(P1.2)}\) is given as follows.
\begin{subequations}
\begin{align*}
\mathrm{(P1.2\text{-}1^\prime)}: \kern-4pt\mathop{\mathtt{max}}_{\mv X, \mv Q_s, \mv w_p, \mv v_p,t,y}\kern-4pt
&~~~\frac{1}{2}c_1\log_2(1+t)+c_2r_{SR}(\mv Q_s)\notag\\
\mathtt {s.t.}& ~~~\eqref{eq:per-EAP power constraint for WIT}\--\eqref{eq:per-EAP power constraint for WPT in terms of X},\ \eqref{eq:constraint for PSD variables and t},\  \eqref{eq:lower-bound SINR of the EAPs superior to that of the ST with fixed rho}\--\eqref{eq:lower-bound SINR of the EAPs larger than t with fixed rho},\  \eqref{eq:maximum permissive interference caused by ST},\ \eqref{eq:affine larger than affine},\ \eqref{eq:sum power constraint with fixed rho}\--\eqref{eq:sum cost constraint with fixed rho},\ \eqref{eq:y larger than zero}.
\end{align*}
\end{subequations}

\subsection{Proposed Iterative Solutions}\label{subsec:Proposed Iterative Solutions}
\textcolor{black}{In this subsection, we propose an iterative algorithm to solve \(\mathrm{(P1)}\) based on investigation into the feasibility of problem \(\mathrm{(P1.1\text{-}1^\prime)}\) and \(\mathrm{((P1.2\text{-}1^\prime)}\).} Due to their similar structure, we focus on studying the feasibility of \(\mathrm{(P1.1\text{-}1^\prime)}\), and then point out some key differences between these two problems in terms of their feasibility. 

First, we solve the following feasibility problem to find a feasible \(\mv X\) to \(\mathrm{(P1.1\text{-}1^\prime)}\).
\begin{subequations}
\begin{align}
\mathrm{(P1.1\text{-}0)}:&~~~\kern-4pt\mbox{Find a solution of $\mv X$ and $\mv v_p$}\kern-4pt\notag\\
\mathtt {s.t.}& ~~~\eqref{eq:per-EAP power constraint for WPT in terms of X},\ \eqref{eq:lower-bound SINR of the ST superior to that of the EAPs with fixed rho},\ \eqref{eq:sum cost constraint with fixed rho},\\
&~~~\mv X\succeq\mv 0.
\end{align}
\end{subequations} Note that it is guaranteed by this step that given \(\bar\varrho\), a feasible \(\mv X\) to problem \(\mathrm{(P1)}\) exists, since for an arbitrary \(\bar{\mv X}\) satisfying \eqref{eq:per-EAP power constraint for WPT in terms of X} and \eqref{eq:sum cost constraint with fixed rho}, either \eqref{eq:lower-bound SINR of the ST superior to that of the EAPs with fixed rho} or \eqref{eq:lower-bound SINR of the EAPs superior to that of the ST with fixed rho} must hold. Therefore, if the \(\bar{\mv X}\) fails to make \(\mathrm{(P1.1)}\) feasible, it must make \(\mathrm{(P1.2)}\) feasible.

Next, applying the returned \(\mv X\) in the above problem as a new initial point of \(\bar{\mv X}\), we aim to find feasible \( t\) and \(y\) by solving the following problem.
\begin{subequations}
\begin{align}
\mathrm{(P1.1\text{-}0^\prime)}:&~~~\kern-4pt\mbox{Find a solution of $\mv X$, $\mv Q_s$, $\mv w_p$, $\mv v_p$, $t$, and $y$}\kern-4pt\notag\\
\mathtt {s.t.}& ~~~\eqref{eq:per-EAP power constraint for WIT}\--\eqref{eq:per-EAP power constraint for WPT in terms of X},\ \eqref{eq:constraint for PSD variables and t},\  \eqref{eq:lower-bound SINR of the ST superior to that of the EAPs with fixed rho}\--\eqref{eq:lower-bound SINR of the ST larger than t with fixed rho},\  \eqref{eq:maximum permissive interference caused by ST},\ \eqref{eq:affine larger than affine},\ \eqref{eq:sum power constraint with fixed rho}\--\eqref{eq:sum cost constraint with fixed rho},\ \eqref{eq:y larger than zero}.
\end{align}
\end{subequations} It is worth noting that proper chosen of \(\bar t\) and \(\bar y\) is necessary to solve \(\mathrm{(P1.1\text{-}0^\prime)}\). For example, with \(\bar{\mv X}\) fixed, \(\bar y\) can be set as \(\sigma_{PR}^2\), and then \(\bar t\) can be set as \(\max\{0,\min\{(1-\bar\varrho)P_pg_1(\bar{\mv X}), \Myfrac{(\gamma^\ast)^2}{y}\}\}\), in which \(\gamma^\ast\) is the optimum value of the following problem.
\begin{align*}
\kern-4pt\mathop{\mathtt{max}}_{\mv w_p, \mv v_p}\kern-4pt
&~~~\Re\left\{\mv g_{sp}^H\mv w_p+\mv g_{EAP,p}^H\mv v_p\right\}\notag\\
\mathtt {s.t.}& ~~~\eqref{eq:per-EAP power constraint for WIT},\ \eqref{eq:sum power constraint with fixed rho},\\
&~~~\|\mv w_p\|^2\mv\le\eta\bar\varrho P_{\rm EH}(\mv X).
\end{align*}
Since problem \(\mathrm{(P1.1\text{-}0)}\) and \(\mathrm{(P1.1\text{-}0^\prime)}\) are both easily observed to be convex problems, they can be optimally solved by some optimization toolboxes such as CVX \cite{CVX}. Denoting \(\mv X\), \(t\) and \(y\) returned by \(\mathrm{(P1.1\text{-}0^\prime)}\) by \(\bar{\mv X}^{(0)}\), \(\bar t^{(0)}\), and \(\bar y^{(0)}\), respectively, it is easily seen that \(\mathrm{(P1.1\text{-}1^\prime)}\) is feasible if \(\mv X\), \(t\), and \(y\) take the value of \(\bar{\mv X}^{(0)}\), \(\bar t^{(0)}\), and \(\bar y^{(0)}\), respectively. Hence, it safely arrives at a feasible problem \(\mathrm{(P1.1\text{-}1^\prime)}\) with the initial points \(\{\bar{\mv X}^{(0)}, \bar t^{(0)}, \bar y^{(0)}\}\) specified as above mentioned.

\begin{remark}
Although the initial points to problem \(\mathrm{(P1.2\text{-}1^\prime)}\) can be found similarly, it is worth noting that compared with \eqref{eq:lower-bound SINR of the EAPs superior to that of the ST with fixed rho}, \eqref{eq:lower-bound SINR of the ST superior to that of the EAPs with fixed rho} turns out to be more likely to be infeasible, since the LHS of \eqref{eq:SINR of the ST superior to that of the EAPs with fixed rho} is a monotonically decreasing function over \(\bar\varrho\in[0,1]\), and when \(\bar\varrho\to 1\), it is hardly lower-bounded by the LHS of \eqref{eq:lower-bound SINR of the ST superior to that of the EAPs with fixed rho}. To illustrate this, we consider the case that there is no \(\mv X\) satisfying \eqref{eq:lower-bound SINR of the ST superior to that of the EAPs with fixed rho} when \(\bar\varrho=1\). Hence, when infeasibility of \(\mathrm{(P1.1\text{-}0)}\) is detected assuming that \(\varrho\) is searched in an increasing order, only \(\mathrm{(P1.2\text{-}1^\prime)}\) needs to be solved for the rest of \(\bar\varrho\).
\end{remark}

Since \(\mathrm{(P1.1\text{-}1^\prime)}\) and \(\mathrm{(P1.2\text{-}1^\prime)}\) have been shown to be convex problems, and at least one of them is guaranteed to be feasible by initializing \(\{\bar{\mv X}^{(0)}, \bar t^{(0)}, \bar y^{(0)}\}\) as discussed above, an SCA-based algorithm is developed to solve \(\mathrm{(P1)}\) as shown in Algorithm~\ref{alg:Algorithm I}. The convergence behaviour of Algorithm~\ref{alg:Algorithm I} is assured by the following proposition.

\begin{proposition}
Monotonic convergence of solutions to problem \(\mathrm{(P1.1\text{-}1^\prime)}\) and \(\mathrm{(P1.2\text{-}1^\prime)}\) in Algorithm~\ref{alg:Algorithm I} is achieved, i.e., \(\Myfrac{1}{2}c_1\log_2(1+t^{(n)})+c_2r_{SR}(\mv Q_s^{(n)})\ge \Myfrac{1}{2}c_1\log_2(1+t^{(n-1)})+c_2r_{SR}(\mv Q_s^{(n-1)})\). Moreover, the converged solutions satisfy all the constraints as well as the Karush-Kuhn-Tucker (KKT) conditions of problem \(\mathrm{(P1.1\text{-}1)}\) and \(\mathrm{(P1.2\text{-}1)}\), respectively. \label{prop:convergence and KKT}
\end{proposition}

\begin{IEEEproof}
Please refer to Appendix~\ref{appendix:proof of convergence and KKT}.
\end{IEEEproof}

\textcolor{black}{Next, we analyse the complexity of Algorithm~\ref{alg:Algorithm I} in terms of counting the arithmetic operations. Since most off-the-shelf convex optimization toolboxes handle the repeatedly encountered SDP using an interior-point algorithm, the worst-case complexity for solving \(\mathrm{(P1)}\) is given by\footnote{For the simplicity of exposition, we assume \(N_{T,1}=\cdots=N_{T,K}=\frac{L}{2}\) and \(N_{R,1}=\cdots=N_{R,K}=\frac{L}{2}\) in this expression.} \(\frac{1}{\beta}\big(L_1\mathcal{O}\left(\max\{K+4,\frac{LK}{2}+1,N\}^4
\max\{\frac{LK}{2}+1,N\}^{1/2}\right )+L_2\mathcal{O}\big(\max\{K+4,\frac{LK}{2},N+1\}^4\)
\(\max\{\frac{LK}{2},N+1\}^{1/2}\big)\big)\log({}\Myfrac{1}{\varepsilon})\) \cite{TomLuo2010SDR}, which comprises two parts, where the former part accounts for the complexity for solving \(\mathrm{(P1.1)}\); the latter part accounts for the complexity for solving \(\mathrm{(P1.2)}\); \(L_1\) and \(L_2\) denote the number of iterations for the SCA in solving \(\mathrm{(P1.1\text{-}1^\prime)}\) and \(\mathrm{(P1.2\text{-}1^\prime)}\), respectively; \(\beta\) controls the step length for one-dimension search over \(\rho\in[0,1]\); and \(\varepsilon\) is determined by the solution accuracy.}

\begin{algorithm}[tp]
\caption{Proposed Algorithm for Solving Problem \(\mathrm{(P1)}\)}\label{alg:Algorithm I}
\begin{algorithmic}[1]
\Require \(\bar\varrho\); \(\mathtt{flag}_{\rm ST}=1\)
\If {both \(\mathrm{(P1.1\text{-}0)}\) and \(\mathrm{(P1.1\text{-}0^\prime)}\) are solvable}
\State \textbf{go to} \ref{procedure:beginning}
\Else
\State \(\mathtt{flag}_{\rm ST}=0\); \(f_1(\bar\varrho)\gets 0\)
\EndIf
\State \(n\gets 0\); initialize \(\{\bar{\mv X},  \bar t,  \bar y\}\)  with \(\{\bar{\mv X}^{(0)},  \bar t^{(0)},  \bar y^{(0)}\}\) returned by problem \(\mathrm{(P1.1\text{-}0^\prime)}\)\label{procedure:beginning}
\Repeat
\State Solve problem \(\mathrm{(P1.1\text{-}1^\prime)}\) to obtain \(\{{\mv X}^{(n+1)},  t^{(n+1)},  y^{(n+1)}\}\)
\State Update \(\{\bar{\mv X}^{(n+1)}, \bar t^{(n+1)}, \bar y^{(n+1)}\}\gets\{{\mv X}^{(n+1)}, t^{(n+1)}, y^{(n+1)}\}\)
\State \(n\gets n+1\)
\Until convergence of the objective value of \(\mathrm{(P1.1\text{-}1^\prime)}\)
\State \(f_1(\bar\varrho)\gets\)  the optimum value of \(\mathrm{(P1.1\text{-}1^\prime)}\)
\State Obtain \(f_1^\ast\) by one-dimension search over \(\varrho\)

\State Solve problem \(\mathrm{(P1.2)}\) using the SCA method similarly to obtain \(f_2^\ast\) by one-dimension search over \(\varrho\)
\Ensure \(f^\ast=\max\{f_1^\ast, f_2^\ast\}\)
\end{algorithmic}
\end{algorithm}

\subsection{Proposed ZF-Based Solutions}\label{subsec:Proposed ZF-Based Solutions}
\textcolor{black}{In this subsection, we develop an insightful suboptimal solution that simplifies the receiving beamforming design of the full-duplex EAPs.} It is seen from \eqref{eq:first-slot SINR at the EAPs} that the incumbent design of \(\mv u_2\) depends on \(\mv X\), which means that there will be some additional central optimization resources induced to compute \(\mv u_2\) after the problem \(\mathrm{(P1)}\) has been solved and the optimum \(\mv X\) has been returned. The broadcast of \(\mv u_2\) causes unfavourable delay particularly when \(N_{R,k}\)'s is large in practice. Hence, we design \(\mv X\) in such a way that the receiving beamforming \(\mv u_2\) can be locally decided.

To do so, let EAPs jointly decode the PT's message regardless of the residual LI. For example, an arbitrary vector align with \(\mv h_{p,EAP}\) is chosen as \(\mv u_2\), i.e., \(\mv u_2=\mu\mv h_{p,EAP}\), \(\mu\in\mathbb{R}\). In this way, the joint decoding can be implemented with the \(k\)th EAP having access only to its local CSI, i.e., \(\mv h_{p,EAP_k}\)'s. Accordingly, the resulting receiving SINR at the EAPs coincides with its maximum, i.e., \(\Myfrac{P_p\|\mv h_{p,EAP}\|^2}{\sigma_{EAP}^2}\),  if and only iff (iff) \(\mv u_2^H\tilde{\mv H}_{TR}\mv X\tilde{\mv H}_{TR}^H\mv u_2=\mv 0\). Combining with \(\mv u_2=\mu\mv h_{p,EAP}\), \(\mv X\) needs to be designed such that \(\mv h_{p,EAP}^H\tilde{\mv H}_{TR}\mv X\tilde{\mv H}_{TR}^H\mv h_{p,EAP}=0\). Defining \(\mv h=\tilde{\mv H}_{TR}^H\mv h_{p,EAP}\) with its normalized vector denoted by \(\bar{\mv h}\), the projection matrix \(\mv P=\mv I-\bar{\mv h}\bar{\mv h}^H\) can be alternatively expressed as \(\mv P=\tilde{\mv U}\tilde{\mv U}^H\) with \(\tilde{\mv U}\in\mathbb{C}^{\sum N_{T,k}\times (\sum N_{T,k}-1)}\) such that \(\bar{\mv h}^H\tilde{\mv U}=\mv 0\) and \(\tilde{\mv U}^H\tilde{\mv U}=\mv I\). The optimum structure of \(\mv X\) is then specified by the following lemma \cite[Lemma 3.1]{Zhang2010BD}.
\begin{lemma}
The ZF-based \(\mv X\) to \(\mathrm{(P1)}\) is given by
\begin{align}
\mv X=\tilde{\mv U}\tilde{\mv X}\tilde{\mv U}^H, \label{eq:optimum X}
\end{align} where \(\tilde{\mv X}\in\mathbb{C}^{(\sum N_{T,k}-1)\times (\sum N_{T,k}-1)}\) is a PSD matrix. \label{lemma:optimum X}
\end{lemma}

Applying Lemma~\ref{lemma:optimum X} to Problem \(\mathrm{(P1.1)}\), Problem \(\mathrm{(P1.1\text{-}1^\prime)}\) can be reduced to
\begin{subequations}
\begin{align}
\mathrm{(P1.1\text{-}1^\prime\text{-}ZF)}: \kern-12pt\mathop{\mathtt{max}}_{\tilde{\mv X}, \mv Q_s, \mv w_p, \mv v_p, t, y}\kern-12pt
&~~~\frac{1}{2}c_1\log_2(1+t)+c_2r_{SR}(\mv Q_s)\notag\\
\mathtt {s.t.}& ~~~\eqref{eq:per-EAP power constraint for WIT},\ \eqref{eq:maximum permissive interference caused by ST},\ \eqref{eq:affine larger than affine},\\
&~~~{\sf Tr}(\tilde{\mv U}^H\mv E_k\tilde{\mv U}\tilde{\mv X})\le P_0, \; \forall k, \label{eq:per-EAP power constraint for ZF-WPT in terms of X}\\
&~~~(1-\bar\varrho)P_p\left(\tilde g_1(\bar{\mv X})+\Re\left\{{\sf Tr}\left(\nabla_{\mv X}\tilde g_1(\bar{\mv X})(\tilde{\mv X}-\bar{\mv X})\right)\right\}\right)\notag\\
&~~~\ge \Myfrac{P_p\|\mv h_{p,EAP}\|^2}{\sigma_{EAP}^2},\\
&~~~(1-\bar\varrho)P_p\left(\tilde g_1(\bar{\mv X})+\Re\left\{{\sf Tr}\left(\nabla_{\mv X}\tilde g_1(\bar{\mv X})(\tilde{\mv X}-\bar{\mv X})\right)\right\}\right)\ge t,\\
&~~~{\sf Tr}(\mv Q_s)+\|\mv w_p\|^2\mv\le\eta\bar\varrho P_{\rm EH}(\tilde{\mv U}\tilde{\mv X}\tilde{\mv U}^H), \label{eq:sum power constraint for ZF}\\
&~~~c_3\eta\bar\varrho{\sf Tr}(\mv H_{EAP,ST}\tilde{\mv U}\tilde{\mv X}\tilde{\mv U}^H\mv H_{EAP,ST}^H)+c_4\vert \mv g_{EAP,p}^H\mv v_p\vert^2\le C, \label{eq:sum cost constraint for ZF}\\
&~~~\tilde{\mv X}\succeq\mv 0, \, \mv Q_s\succeq\mv 0, \, t\ge 0, \, y\ge 0, \label{eq:constraint for PSD in the ZF case, t, and y}
\end{align}
\end{subequations} where\(\tilde g_1(\mv X)=g_1(\tilde{\mv U}\tilde{\mv X}\tilde{\mv U}^H)\), and \(\nabla_{\mv X}\tilde g_1(\mv X)\) denotes the gradient matrix of \(\tilde g_1(\mv X)\) expressed as \(\nabla_{\mv X}\tilde g_1(\mv X)=\)
\begin{align}
-(1-\bar\varrho)\tilde{\mv U}^H\mv H_{EAP,ST}^H{\mv A}^{-1}(\bar\varrho, \tilde{\mv U}\tilde{\mv X}\tilde{\mv U}^H)\mv h_{p,ST}\mv h_{p,ST}^H{\mv A}^{-1}(\bar\varrho, \tilde{\mv U}\tilde{\mv X}\tilde{\mv U}^H)\mv H_{EAP,ST}\tilde{\mv U}.
\end{align}

Similarly, Problem \(\mathrm{(P1.2\text{-}1^\prime)}\) reduces to the following problem.
\begin{subequations}
\begin{align}
\mathrm{(P1.2\text{-}1^\prime\text{-}ZF)}: \kern-12pt\mathop{\mathtt{max}}_{\tilde{\mv X}, \mv Q_s, \mv w_p, \mv v_p,t,y}\kern-12pt
&~~~\frac{1}{2}c_1\log_2(1+t)+c_2r_{SR}(\mv Q_s)\notag\\
\mathtt {s.t.}& ~~~\eqref{eq:per-EAP power constraint for WIT}, \eqref{eq:per-EAP power constraint for ZF-WPT in terms of X},\  \eqref{eq:maximum permissive interference caused by ST},\ \eqref{eq:affine larger than affine},\ \eqref{eq:sum power constraint for ZF}\--\eqref{eq:constraint for PSD in the ZF case, t, and y},\\
&~~~\frac{P_p\|\mv h_{p,EAP}\|^2}{\sigma_{EAP}^2}\ge (1-\bar\varrho)P_p\mv h_{p,ST}^H\mv A^{-1}(\bar\varrho, \tilde{\mv U}\tilde{\mv X}\tilde{\mv U}^H)\mv h_{p,ST},\label{eq:SINR of the ZF EAPs superior to that of the ST}\\
&~~~\frac{P_p\|\mv h_{p,EAP}\|^2}{\sigma_{EAP}^2}\ge t.\label{eq:SINR of the ZF EAPs larger than t}
\end{align}
\end{subequations}

Note that compared with \(\mathrm{(P1.2\text{-}1^\prime)}\), \(\mathrm{(P1.2\text{-}1^\prime\text{-}ZF)}\) admits a substantially simplified exposition because not only is there no more variable related to \(\mv X\) in the LHS of \eqref{eq:SINR of the ZF EAPs superior to that of the ST} and \eqref{eq:SINR of the ZF EAPs larger than t}, but also they turn out to be convex. This means that there is no approximation made w.r.t \(\mv X\), and thus \(\mathrm{(P1.2\text{-}1^\prime\text{-}ZF)}\) is expected to converge faster, since there are only two iterated variables remained, \(t\) and \(y\). As the ZF-based solutions are reduced from Problem \(\mathrm{(P1)}\), the solution and the convergence analysis are thus similar to Algorithm~\ref{alg:Algorithm I}. Hence, we only present an outline of the algorithm for the proposed ZF-based solutions which is shown in Algorithm~\ref{alg:Algorithm II}. \textcolor{black}{The worst-case complexity using the ZF-based solutions can be analysed in analogue to that using the proposed iterative solutions, which is given by \(\frac{1}{\beta}\big(L_1^\prime\mathcal{O}\left(\max\{K+5,\frac{LK}{2}-1,N\}^4\max\{\frac{LK}{2}-1,N\}^{1/2}\right )\) \(+ L_2^\prime\mathcal{O}\big(\max\{K+3,\frac{LK}{2}-1,N+1\}^4\) \(\max\{\frac{LK}{2}-1,N+1\}^{1/2}\big)\big)\log({}\Myfrac{1}{\varepsilon})\), where \(L_1^\prime\) and \(L_2^\prime\) denote the number of iterations for the SCA in solving \(\mathrm{(P1.1\text{-}1^\prime\text{-}ZF)}\) and \(\mathrm{(P1.2\text{-}1^\prime\text{-}ZF)}\), respectively.} 

\begin{algorithm}[tp]
\caption{Proposed ZF-Based Algorithm for Solving Problem \(\mathrm{(P1)}\)}\label{alg:Algorithm II}
\begin{algorithmic}[1]
\State Find feasible \(\{\tilde{\mv X}, t, y\}\) to \(\mathrm{(P1.1\text{-}1^\prime\text{-}ZF)}\) as the initial \(\{\bar{\mv X}^{(0)}, \bar t^{(0)}, \bar y^{(0)}\}\)
\State Solve \(\mathrm{(P1.1\text{-}1^\prime\text{-}ZF)}\) using the SCA method
\State Solve \(\mathrm{(P1.1)}\) to obtain \(f_1^\ast\) by one-dimension search over \(\varrho\)
\State Find feasible \(\{t, y\}\) to \(\mathrm{(P1.2\text{-}1^\prime\text{-}ZF)}\) as the initial \(\{\bar t^{(0)}, \bar y^{(0)}\}\)
\State Solve \(\mathrm{(P1.2\text{-}1^\prime\text{-}ZF)}\) using the SCA method
\State Solve \(\mathrm{(P1.2)}\) to obtain \(f_2^\ast\) by one-dimension search over \(\varrho\)
\Ensure \(f^\ast=\max\{f_1^\ast, f_2^\ast\}\)
\end{algorithmic}
\end{algorithm}

\section{Benchmark Schemes}\label{sec:Benchmark Schemes}
In this section, two benchmark schemes are presented, where only one of the available EAPs operating in FD mode is selected to assist with the CCRN, and all the EAPs work together but operate in HD mode, respectively.

\subsection{Selective Non-cooperative FD Scheme}\label{subsec:FD Non-cooperative}
First, consider the case when only one EAP is selected in the CCRN. This is the case when joint transmission and/or detection of the WPT and WIT signals is expensive or unavailable due to extra resources (e.g., spectrum, power, centralized coordination point) or strict synchronization requirement among EAPs. The selection of the EAP is based on a simple criterion: \(\tilde k=\arg\max\limits_{k\in\mathcal{K}}\|\mv h_{p,EAP_k}\|^2\). Note that by replacing \(\mv h_{p,EAP}\) (\(\mv g_{EAP,p}\)) with \(\mv h_{p,EAP_{\tilde k}}\) (\(\mv g_{EAP_{\tilde k},p}\)), solving the resultant \(\mathrm{(P1)}\) follows the same procedure as those detailed in Section~\ref{subsec:Proposed Iterative Solutions}, and thus is omitted here for brevity. \textcolor{black}{Note that the worst-case complexity for solving \(\mathrm{(P1)}\) based on the selective non-cooperative solutions can be attained by simply substituting \(K\) by \(1\) in the complexity analysis for the proposed iterative solutions.}

\subsection{HD EB and DF Relaying}\label{subsec:HD}
Next, consider the case when all the EAPs work under the HD mode. In this case, as DF relaying only takes place at the ST, \(r_{PR}(\mv X, \varrho)\) reduces to
\begin{align}
r_{PR}^\prime(\mv X,\varrho)=\frac{1}{2}\log_2\left(1+\min\left\{(1-\varrho)P_p\mv h_{p,ST}^H\mv A^{-1}\mv h_{p,ST}, \ \frac{\left\vert\mv g_{sp}^H\mv w_p\right\vert^2}{\mv g_{sp}^H\mv Q_s\mv g_{sp}+\sigma_{PR}^2}\right\}\right).
\end{align}
Moreover, the total cost constraint for the HD EAPs-aided CCRN is also simplified as
\begin{align}
c_3\eta\varrho{\sf Tr}(\mv H_{EAP,ST}\mv X\mv H_{EAP,ST}^H)\le C. \label{eq:sum cost constraint for HD}
\end{align}
Accordingly, \(\mathrm{(P1)}\) can be equivalently recast into the following problem.
\begin{align}
\mathrm{(P1\text{-}HD)}: \kern-4pt\mathop{\mathtt{max}}_{\mv X, \mv Q_s, \mv w_p, \varrho, t}\kern-4pt
&~~~\frac{1}{2}c_1\log_2(1+t)+c_2r_{SR}(\mv Q_s)\notag\\
\mathtt {s.t.}& ~~~\eqref{eq:per-EAP power constraint for WPT in terms of X}\--\eqref{eq:sum power constraint},\ \eqref{eq:SINR of the ST larger than t},\ \eqref{eq:constraint for varrho}\--\eqref{eq:constraint for PSD variables and t},\ \eqref{eq:sum cost constraint for HD}, \\
&~~~\frac{\left\vert\mv g_{sp}^H\mv w_p\right\vert^2}{\mv g_{sp}^H\mv Q_s\mv g_{sp}+\sigma_{PR}^2}\ge t. \label{eq:minimum PR's SINR threshold in the HD case}
\end{align}

Compared with solving \(\mathrm{(P1)}\), we use a slightly different approach to solve \(\mathrm{(P1\text{-}HD)}\) in view of its  structure. Note from \(\mathrm{(P1\text{-}HD)}\) that as \(t\) increases, the objective value of \(\mathrm{(P1\text{-}HD)}\) becomes larger at first. However, continuously increasing \(t\) will eventually violate \eqref{eq:SINR of the ST larger than t} and/or \eqref{eq:minimum PR's SINR threshold in the HD case}, since the LHS of both of them are easily shown to be upper-bounded. On the other hand, it is observed that a large enough \(t\) obtained by suppressing the value of \({\sf Tr}(\mv Q_s)\) may also compromise the value of \(r_{SR}(\mv Q_s)\). Hence, it suggests that there exits a proper \(t\) that achieves the optimum value of \(\mathrm{(P1\text{-}HD)}\). Given \(\varrho=\bar\varrho\) and \(t=\bar t\), denote the optimum value of the following problem  by \(f_1^\prime(\bar\varrho, \bar t)\).
\begin{align}
\mathrm{(P1\text{-}HD\text{-}SDR)}: \kern-4pt\mathop{\mathtt{max}}_{\mv X, \mv Q_s, \mv W_p}\kern-4pt
&~~~\frac{1}{2}c_1\log_2(1+\bar t)+c_2r_{SR}(\mv Q_s)\notag\\
\mathtt {s.t.}& ~~~\eqref{eq:per-EAP power constraint for WPT in terms of X},\ \eqref{eq:lower-bound SINR of the ST larger than t with fixed rho},\\
&~~~{\sf Tr}(\mv G_{sp}\mv W_p)\ge \bar t\left(\mv g_{sp}^H\mv Q_s\mv g_{sp}+\sigma_{PR}^2\right), \\
&~~~{\sf Tr}(\mv Q_s+\mv W_p)\le\eta\bar\varrho P_{EH}(\mv X),\\
&~~~c_3\eta\bar\varrho{\sf Tr}(\mv H_{EAP,ST}\mv X\mv H_{EAP,ST}^H)\le C,\\
&~~~\mv X\succeq\mv 0, \, \mv Q_s\succeq\mv 0, \, \mv W_p\succeq\mv 0,
\end{align} where \(\mv G_{sp}=\mv g_{sp}\mv g_{sp}^H\), and \(\mv W_p=\mv w_p\mv w_p^H\) with its constraint \({\rm rank}(\mv W_p)=1\) removed. Then we have the following lemma \cite[Appendix A]{xing2015multi-AFarXiv}.
\begin{lemma}
\(f_1^\prime(\varrho,t)\) is a concave function w.r.t \(t\).  \label{lemma:obj function concave w.r.t t}
\end{lemma}

Note that given \(\varrho=\bar\varrho\) and \(t=\bar t\), \eqref{eq:lower-bound SINR of the ST larger than t with fixed rho} implies \eqref{eq:SINR of the ST larger than t}, and therefore leads to a lower-bound solution to \(\mathrm{(P1\text{-}HD)}\), while imposing rank relaxation on \(\mv W_p\) in general enlarges its feasible region and thus yields an upper-bound solution to \(\mathrm{(P1\text{-}HD)}\). The final effect of these two transformation on the objective value of problem \(\mathrm{(P1\text{-}HD)}\) is nevertheless of no ambiguity, since the tightness of the rank relaxation in the latter holds because of the following proposition.
\begin{proposition}
The optimal solution to problem \(\mathrm{(P1\text{-}HD\text{-}SDR)}\) satisfies \({\rm rank}(\mv W_p^\ast)=1\) such that \(\mv W_p^\ast=\mv w_p^\ast\mv w_p^{\ast H}\). \label{prop:rank one of W_p}
\end{proposition}
\begin{IEEEproof}
The proof for the rank-one property of \(\mv W^\ast\) is similar to \cite[Appendix A]{xing2016GC}, and is thus omitted here due to the length constraint of the paper.
\end{IEEEproof}
As a result, we solve \(\mathrm{(P1\text{-}HD)}\) by two-dimension search over \(\varrho\) and \(t\), i.e., \(f^{\prime\ast}=\max\limits_{\varrho,t}f_1^\prime(\varrho,t)\), where \(t\) can be found by some low-complexity search such as bi-section algorithm in accordance with Lemma~\ref{lemma:obj function concave w.r.t t}. Specifically, given any \(\varrho\) and \(t\), an SDR problem shown in \(\mathrm{(P1\text{-}HD\text{-}SDR)}\) is solved with only one SCA-approximated constraint \eqref{eq:lower-bound SINR of the ST larger than t with fixed rho}. \textcolor{black}{The worst-case complexity for solving \(\mathrm{(P1\text{-}HD)}\) is accordingly given by \(\frac{L_3}{\beta}\mathcal{O}\left(\max\{K+4,\frac{LK}{2},N\}^4\max\{\frac{LK}{2},N\}^{1/2}\right )\log({}\Myfrac{t_{\max}}{\beta_{\rm bi}})\) \(\log({}\Myfrac{1}{\varepsilon})\), where \(L_3\) denotes the number of iterations for the SCA in solving \(\mathrm{(P1\text{-}HD\text{-}SDR)}\) and \(\beta_{\rm bi}\) represents the step length for bi-section w.r.t \(t\).}

\section{Numerical Results}\label{sec:Numerical Results}
In this section, we evaluate the proposed joint EB and DF relaying scheme aided by multiple FD-enabled EAPs in the CCRN against the benchmark schemes. The proposed iterative solutions and the ZF-based solutions for solving \(\mathrm{(P1)}\) in Section~\ref{subsec:Proposed Iterative Solutions} and Section~\ref{subsec:Proposed ZF-Based Solutions} are denoted by ``FD Proposed'' and ``FD ZF'', respectively. For the benchmarks, the non-cooperative scheme with only one EAP associated with the ST in Section~\ref{subsec:FD Non-cooperative} is denoted by ``FD Non-cooperative'', while the HD case introduced in Section~\ref{subsec:HD} is denoted by ``HD''.

In the following numerical examples, the parameters are set as follows unless otherwise specified. As illustrated in Fig.~\ref{fig:transmission protocol}, there is one PT, one PR, each with one single antenna, and a pair of multi-antenna ST and SR equipped with \(M=2\) and \(N=2\) antennas, respectively. There are also \(K=3\) EAPs each equipped with \(L=4\) antennas, among which half of them are specified as transmitting antennas and the other half as receiving antennas, i.e., \(N_{T,1}=N_{T,2}=N_{T,3}=2\) and \(N_{R,1}=N_{R,2}=N_{R,3}=2\). The distance from the ST to the PT, PR, and SR are set as \(d_{p,ST}=10\)m, \(d_{sp}=10\)m and \(d_{ss}=10\)m.
The EAPs are located within a circle centred on the ST with their radius uniformly distributed over \([0,10]\)m. The generated wireless channels consist of both large-scale path loss and small-scale multi-path fading. The pathloss model is given by \(A_0(\Myfrac{d}{d_0})^{-\alpha}\) with \(A_0=-30\)dB , where \(d\) denotes the relevant distance, \(d_0=1\)m is a reference distance, and \(\alpha=2.5\) is the path loss exponent factor. The small-scale fading follows $i.i.d.$ Rayleigh fading with zero mean and unit variance. The effective residual LI channel gain \(\theta^2\) is set to be \(-60\)dB.  The weight coefficients are assumed to be \(c_1=c_2=1\). \(c_3=1\) and \(c_4=1\) are adopted for the cost per unit of received energy for WPT and WIT, respectively. The normalized cost, defined by \(\Myfrac{C}{C_{\max}}\), is set to be \(0.1\). The transmitting power is set as \(P_p=10\)dBm and \(P_0=20\)dBm.  The other parameters are set as follows. The RF-band AWGN noise and the RF-band to baseband conversion noise are set to be \(\sigma_{n_a}^2=-110\)dBm and \(\sigma_{n_c}^2=-70\)dBm, respectively; \(\sigma_{EAP}^2=\sigma_{PR}^2=\sigma_{SR}^2\) are all set equal to \(\sigma_{n_a}^2+\sigma_{n_c}^2\); and the EH efficiency is set as \(\eta=50\%\) \cite{nasir16multicell}. The evaluation in the following examples are averaged over $300$ independent channel realizations.

\begin{figure}[htp]
\begin{center}
\includegraphics[width=3.5in]{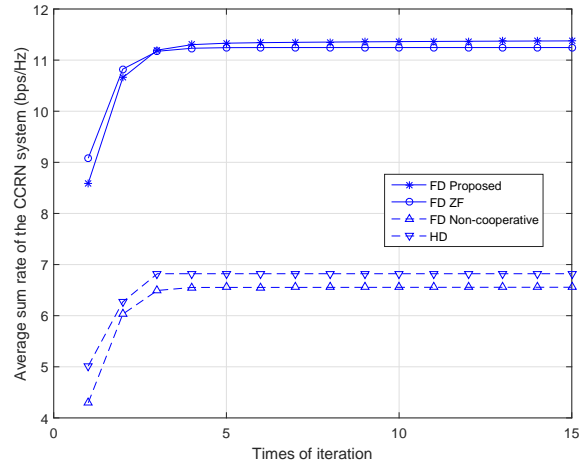}
 \end{center}
\caption{The average sum-rate of the CCRN system versus the times of iteration of the SCA algorithm by the four schemes, in which \(c_1=c_3=c_4=1\) and \(c_2=10\).}\label{fig:sum rate vs Iter}
\vspace{-0.2in}
\end{figure}
Fig.~\ref{fig:sum rate vs Iter} shows the convergence behaviour of the proposed iterative algorithm, which is guaranteed by Proposition~\ref{prop:convergence and KKT}. It is also seen that the number of iterations for the proposed and the suboptimal schemes  to converge is around within \(10\) , while that for the benchmark schemes `FD Non-cooperative' and `HD' are less than \(5\). It is also observed that there is little performance gap between ``FD ZF'' solutions and ``FD Proposed'' solutions.

\begin{figure}[!htp]
\begin{center}
\includegraphics[width=3.5in]{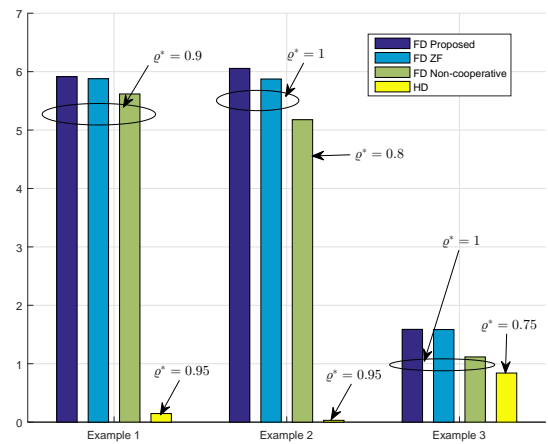}
 \end{center}
\caption{The instantaneous sum-rate of the CCRN system achieved by different schemes in special scenarios, in which \(P_0=23\)dBm and \(P_1=20\)dBm.}\label{fig:examples}
\vspace{-0.2in}
\end{figure}
Fig.~\ref{fig:examples} shows the instantaneous sum-rate of the system achieved by different schemes and the associated values of the PS factor \(\varrho\) in some special cases. In Example~1, There are \(K=2\) EAPs located on the right of the ST alongside the PT-ST direction, with \(10\)m and \(20\)m away from the ST, respectively; \(d_{p,ST}=5\)m; \(N=4\); \(\theta^2=-40\)dB; \(c_3=10\); and the normalized cost constraint is \(0.01\). The optimal \(\varrho\) for all the cases except the ``HD'' is about \(0.9\), which means that the ST does not exploit its full EH capability. This is mainly due to the following two reasons. On one hand, since \(\mv h_{p,ST}\) is better than \(\mv h_{p,EAP}\), the optimum value of \(f\) is achieved by \(f_1^\ast\), which  means that the constraint in \eqref{eq:SINR of the ST larger than t} is active. As a result, continuing increasing \(\varrho\) will violate \eqref{eq:SINR of the ST larger than t}, since it is not hard to prove that the LHS of is a monotonically decreasing function w.r.t \(\varrho\in[0,1)\). On the other hand, since \(c_3\) is as ten times large as \(c_4\), which means that the unit price required by WPT is quite high, the system intuitively prefers to saving the amount of harvested power in this case. In Example~2, there are \(K=3\) EAPs uniformly located on a circle of radius \(10\)m centred on the ST, \(d_{p,ST}=10\)m, and all the other settings are the same as those in Example~1. It is seen that the optimal \(\varrho\) is one in this case for both of the ``FD Proposed'' and ``FD ZF'' scheme, which is apparently due to the good channel condition of \(\mv h_{p,EAP}\), and the enhanced \(\mv H_{EAP,ST}\), which motivates this condition favourable to WPT.

Example~3 explores the other special case when ``HD'' scheme also performs reasonably well. In this case there are \(K=5\) EAPs uniformly distributed on a circle of radius \(5\)m centred on the ST; \(d_{ss}=5\)m; \(L=2\), \(c_1=0.1\), \(c_2=1.9\), and \(c_4=100\); and the normalized cost constraint is \(0.001\). Note that this case emulates the scenario when the weighted sum-rate imposes priority on the secondary system subject to a quite limited cost budget. Since the secondary system's rate is contributed by the ST's harvested power, the priority in favour of the ST's transmission is achieved by splitting all its received power for EH while leaving the task of relaying the PT's message to the EAPs.

\begin{figure}[htp]
\centering
\subfigure[\(\Myfrac{C}{C_{\max}}=0.1\)\label{subfig:nc ten percent}]{\includegraphics[width=3.20in]{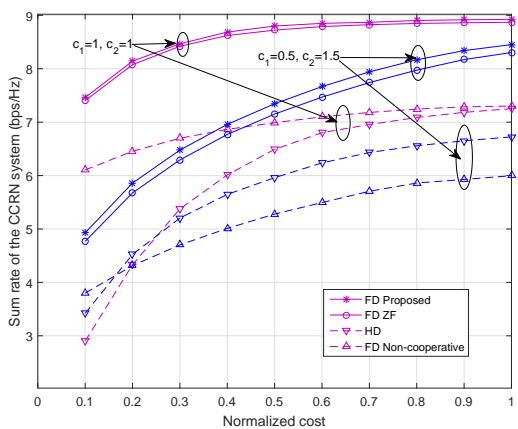}}
\subfigure[\(\Myfrac{C}{C_{\max}}=0.01\)\label{subfig:nc one percent}]{\includegraphics[width=3.20in]{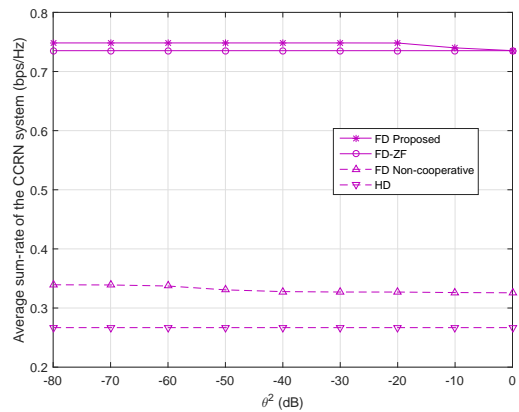}}
\caption{The average sum-rate of the CCRN system versus the residue power level of the LI, in which \(d_{p,ST}=15\)m, \(d_{sp}=10\)m, \(d_{ss}=10\)m, \(c_1=0.1\), \(c_2=1.9\), \(c_3=0.1\), and \(c_4=10\); the \(K=3\) EAPs are located within a ST-centred circle with their radius uniformly distributed over \([0,5]\)m.}\label{fig:sum rate vs theta}
\vspace{-0.2in}
\end{figure}
Fig. ~\ref{fig:sum rate vs theta} reflects the impact of the residue power of LI on the average sum-rate of the system subject to different normalized cost constraints by different schemes. It is observed that the FD schemes are in general quite robust against the increasing power of LI. It is also seen that the suboptimal ``FD ZF'' approaches ``FD Proposed'' with negligible gap when \(\theta^2\) is larger than \(-20\)dB. \textcolor{black}{This is because ``FD Proposed'' solutions tend to substantially suppress the LI received by the EAPs such that \(\mv u_2^{\ast H}\tilde{\mv H}_{TR}\mv X^\ast\tilde{\mv H}_{TR}^H\mv u_2^\ast\approx 0\) (c.f.~\eqref{eq:first-slot SINR at the EAPs}), and therefore it is overall less affected by \(\theta^2\).} In addition, ``FD ZF'' and ``HD'' schemes remain exactly the same, since their designs are irrelevant to \(\theta^2\). Moreover, it is intriguing to see that ``HD'' considerably outperforms ``FD Non-cooperative'' in Fig.~\ref{subfig:nc ten percent}, and reversely performs in Fig.~\ref{subfig:nc one percent}. This can be explained as follows. As a result of the transmission priority imposed on the secondary system (\(c_1=0.1\), \(c_2=1.9\)), as well as the relatively cheaper per unit price for WPT (\(c_3=0.1\), \(c_4=10\)), the system tends to power the ST for improving on its own transmission as much as possible. Hence, in this case, the WPT capability of ``HD'' is larger than that of ``FD Non-cooperative'', which leads to a larger objective value that is dominated by the SR's contribution in this case.

\begin{figure}[htp]
\begin{center}
\includegraphics[width=3.5in]{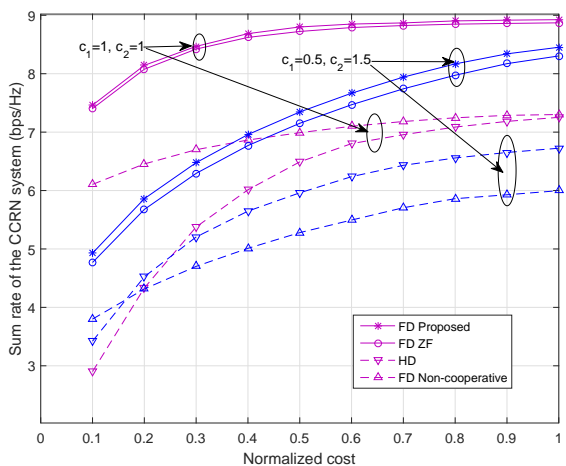}
 \end{center}
\caption{The average weighted sum-rate of the CCRN system versus the normalized cost budget, in which \(K=2\), \(N=6\), \(c_1=0.5\), and \(c_2=1.5\).}\label{fig:sum rate vs cost index}
\vspace{-0.20in}
\end{figure}
Fig.~\ref{fig:sum rate vs cost index} illustrates the average sum-rate of the system achieved by different schemes versus the normalized cost constraints with different weights of the sum-rate. It is observed that the average sum-rate of the system goes up drastically when the transmission is in favour of the ST, which is mainly caused by the imbalanced transmission efficiency between WPT and WIT.
It is also observed that ``FD ZF'' performs nearly as well as ``FD Proposed'' when the primary and the secondary system share the same weights of sum-rate. This is because when \(r_{SR}(\mv Q_s)\) contributes more to the weighted sum-rate, the requirement of increasing \({\sf Tr}(\mv Q_s)\) leads to the fact that the WPT plays a more important role in the CCRN, and therefore the suboptimal design of the WPT transmission will compromise the objective value.

\begin{figure}[htp]
\begin{center}
\includegraphics[width=3.5in]{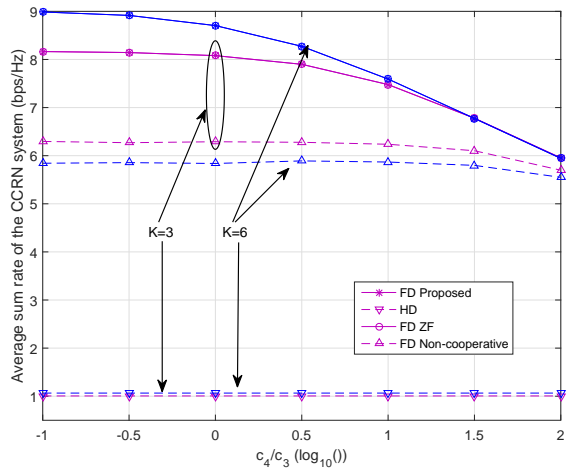}
 \end{center}
\caption{The average sum-rate of the CCRN system versus the unit price of WIT normalized by that of WPT, in which \(P_p=20\)dBm and there is a constant total cost \(C\) set to be \(3\).}\label{fig:sum rate vs cost ratio}
\vspace{-0.2in}
\end{figure}
The way that the unit price of the received energy for WIT versus WPT affects the average sum-rate of the system is shown in Fig.~\ref{fig:sum rate vs cost ratio}. It is seen that ``FD ZF'' approaches the proposed ``FD Proposed'' scheme with negligible gap, and both of them fall over the increasing per-unit cost for WIT, with the superior scheme of \(K=6\) eventually decreasing to nearly the same value as that for \(K=3\). These observations are particularly useful when the cost for WIT is higher than that for WPT, which is  usually the case in practice, since compared to coordinated WPT relying on random energy beams, it costs the EAPs more to perform WIT. In addition, ``FD Non-cooperative'' with \(K=6\) EAPs is outperformed by that with \(K=3\) EAPs, which reveals  that the \(\max\limits_{k\in\mathcal{K}}\|\mv h_{p,EAP_k}\|^2\)-based non-cooperative scheme is not an optimal strategy to fully exploit the diversity gains, since it only benefits the first hop of the DF relaying. In other words, the EAP with \(\max\limits_{k\in\mathcal{K}}\|\mv h_{p,EAP_k}\|^2\) does not necessarily possess the maximum  \(\|\mv g_{EAP_k,p}\|^2\) (c.f.~\eqref{eq:SINR at the PR}).

\begin{figure}[htp]
\begin{center}
\includegraphics[width=3.5in]{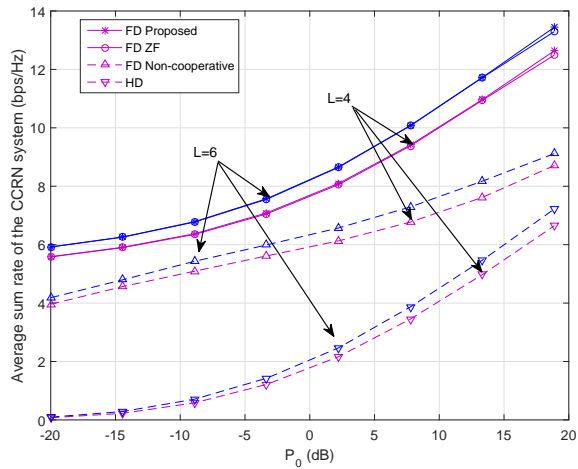}
 \end{center}
\caption{The average sum-rate of the CCRN system versus the per-EAP transmit power constraint, in which \(L=6\).}\label{fig:sum rate vs P0}
\vspace{-0.2in}
\end{figure}
The benefit of increasing per-EAP transmission power for the average sum-rate of the system achieved by different schemes is shown in Fig.~\ref{fig:sum rate vs P0}. It is seen that with larger number of antennas equipped at each EAP, better performance is achieved due to the increasing array gain. It is also noticed that ``FD ZF'' keeps up with ``FD Proposed'' with negligible gap until \(P_0\) increases to \(20\)dB. Moreover, for both cases of \(L=4\) and \(L=6\), it is observed that the average sum-rate achieved by all the schemes other than ``FD Non-cooperative'' goes up quickly as a result of the substantially enlarged feasible region (c.f.~\eqref{eq:per-EAP power constraint for WIT}\--\eqref{eq:per-EAP power constraint for WPT in terms of X}). 

\begin{figure}[htp]
\begin{center}
\includegraphics[width=3.5in]{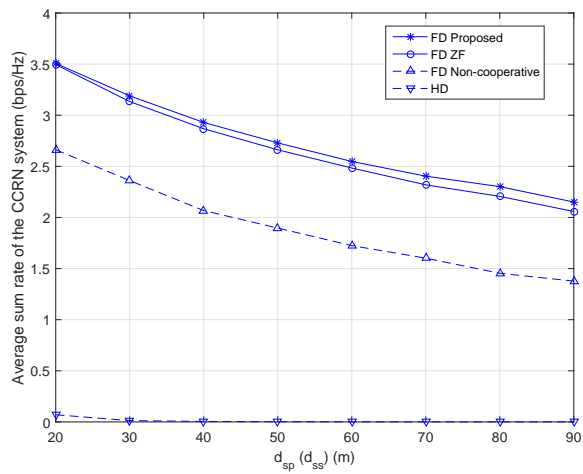}
 \end{center}
\caption{The average weighted sum-rate of the CCRN system versus the WIT distsance between the ST and the SR(PR), in which \(L=6\), \(P0=20dB\)m, \(c_1=0.5\) and \(c_2=1.5\).}\label{fig:sum rate vs distance}
\vspace{-0.2in}
\end{figure}
\textcolor{black}{Considering different sensitivities w.r.t received power for WPT and WIT, given fixed distance from the PT to the ST (\(d_{p,ST}=10\)m), the impact of varying WIT distance (assuming \(d_{sp}=d_{ss}\)) on the average weighted sum-rate of the CCRN system is shown in Fig.~\ref{fig:sum rate vs distance}. It is seen that all of the schemes fall over the WIT distance, and in particular, when the Rxs are more than \(30\)m away from the ST, ``HD'' cannot support effective cooperative transmission any more due to the limited harvested power at the ST.}

\begin{figure}[htp]
\begin{center}
\includegraphics[width=3.5in]{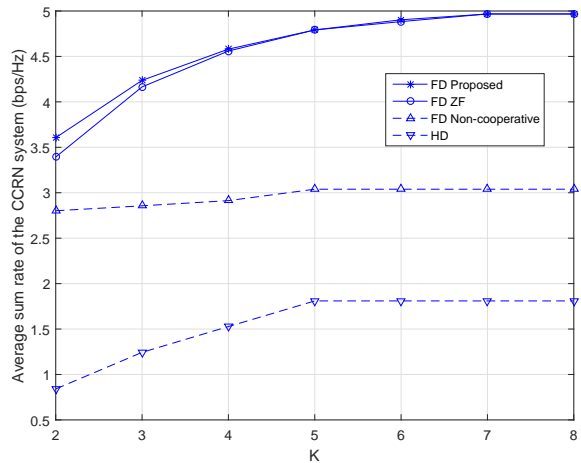}
\end{center}
\caption{The average weighted sum-rate of the CCRN system versus the number of EAPs, in which \(C=4\), \(c_1=0.5\) and \(c_2=1.5\).}\label{fig:sum rate vs number of EAPs}
\vspace{-0.2in}
\end{figure}
\textcolor{black}{The performance of different schemes under different number of FD-enabled EAPs is studied in Fig.~\ref{fig:sum rate vs number of EAPs}, in which a set of EAPs with their distance to the ST drawn from uniform distribution over \([0,10]\)m are first deployed and then allowed to connect to the ST with an increment of one EAP each time. It is observed that the advantage of cooperative gain brought by more involved EAPs is more obviously seen in the cooperative schemes than in those non-cooperative ones.}

\section{Conclusion}\label{sec:Conclusion}
This paper investigated two techniques to fundamentally improve the spectrum efficiency of the RF EH-enabled CCRN, namely, dedicated EB and FD relaying both provided by multi-antenna EAPs. Specifically, assuming  a two-equal-slot DF relaying protocol, the EAPs jointly transfer wireless power to the ST while decoding PT's message in the first transmission phase, the EAPs cooperate to forward PT's message and the ST superimposes PT's message on its own to broadcast in the second transmission phase. The EAPs' EBs as well as their receiving and transmitting beamforming for PT's message, and ST's PS ratio as well as its transmitting beamforming were jointly optimized to maximize the weighted sum-rate taking both energy and cost constraints into account. The proposed algorithms using SCA techniques were proved to converge with the KKT conditions satisfied. The EB design based on ZF was also shown to be promising. Other benchmark schemes were also provided to validate the effectiveness of the proposed ones.

\appendices
\section{Convexity of \eqref{eq:SINR of the ST superior to that of the EAPs with fixed rho}}\label{appendix:proof of convexity of the constraint of ST's receiving SINR superior to that of the EAPs}
First, the gradient of \(g_1(\mv X)\) w.r.t \(\mv X\) is expressed as
\begin{align}
\nabla_{\mv X}g_1(\mv X)=-(1-\bar\varrho)\mv H_{EAP,ST}^H\mv A^{-1}(\bar\varrho,\mv X)\mv h_{p,ST}\mv h_{p,ST}^H\mv A^{-1}(\bar\varrho,\mv X)\mv H_{EAP,ST}. \label{eq:gradient of g_1}
\end{align}
Before obtaining the Hessian matrix of \(g_1(\mv X)\) w.r.t \(\mv X\), we derive the derivative matrix of \eqref{eq:gradient of g_1} as follows. \(D\left(\nabla_{\mv X}g_1(\mv X) \right )=\)
\begin{multline}
(1-\bar\varrho)\mv H_{EAP,ST}^H\mv A^{-1}(\bar\varrho,\mv X)D\left(\mv A(\bar\varrho,\mv X) \right )\mv A^{-1}(\bar\varrho,\mv X)\mv h_{p,ST}\mv h_{p,ST}^H\mv A^{-1}(\bar\varrho,\mv X)\mv H_{EAP,ST}\\
+(1-\bar\varrho)\mv H_{EAP,ST}^H\mv A^{-1}(\bar\varrho,\mv X)\mv h_{p,ST}\mv h_{p,ST}^H\mv A^{-1}(\bar\varrho,\mv X)D\left( \mv A(\bar\varrho,\mv X)\right )A^{-1}(\bar\varrho,\mv X)\mv H_{EAP,ST}. \label{eq:derivative of gradient of g_1}
\end{multline}
Next, in line with the equality \(D\left( \mv A(\bar\varrho,\mv X)\right )=(1-\bar\varrho)\mv H_{EAP,ST}D\mv X \mv H_{EAP,ST}^H\), it follows that
\begin{align}
\nabla_{\mv X}^2g_1(\mv X)=(1-\bar\varrho)^2\left(\mv A_1^T(\bar\varrho,\mv X)\otimes\mv A_2(\bar\varrho,\mv X)+\mv A_2^T(\bar\varrho,\mv X)\otimes\mv A_1(\bar\varrho,\mv X)\right),
\end{align}
where \(\mv A_1(\bar\varrho,\mv X)\) is given by
\begin{align}
\mv A_1(\bar\varrho,\mv X)=\mv H_{EAP,ST}^H\mv A^{-1}(\bar\varrho,\mv X)\mv h_{p,ST}\mv h_{p,ST}^H\mv A^{-1}(\bar\varrho,\mv X)\mv H_{EAP,ST},
\end{align} and \(\mv A_2=\mv H_{EAP,ST}^H\mv A^{-1}(\bar\varrho,\mv X)\mv H_{EAP,ST}\).

We can now determine the convexity of \(g_1(\mv X)\) by studying the semidefiniteness of \(\nabla_{\mv X}^2g_1(\mv X)\) \cite{boyd2004convex}. Take \(\mv A_1^T\otimes\mv A_2\) as an example, Since \(\lambda_l(\mv A_1^T\otimes\mv A_2)=\lambda_l(\mv A_1^T)\lambda_l(\mv A_2)\ge 0\) \cite{horn2012matrix}, where \(\lambda_l(\cdot)\) denotes the \(l\)th non-zero eigenvalue of the associate matrix (\(l=1\) herein), it turns out that \(\mv A_1^T\otimes\mv A_2\) is a PSD matrix and so is \(\mv A_2^T\otimes\mv A_1\). Hence \(\nabla_{\mv X}^2g_1(\mv X)\) is proved to be PSD and so is \(\nabla_{\mv X}^2g_2(\mv X)\),  which completes the proof.

\section{Proof of Lemma~\ref{lemma:the same direction of two inner-products}}\label{appendix:proof of the same direction of two inner-produts}
This can be proved by contradiction. Assuming the optimum value of problem \(\mathrm{(P1.1)}\) is achieved by \(\mv X^\ast\), \(\mv Q^\ast\), \(\mv w_p^\ast\), \(\mv v_p^\ast\), \(t^\ast\) and \(\varrho^\ast\) such that \(\measuredangle(\mv g_{sp}^H\mv w_p^\ast)\neq\measuredangle(\mv g_{EAP,p}^H\mv v_p^\ast)\). In other words, \(\exists\mv w_p^\prime=\mv w_p^\ast \exp\{j\measuredangle\theta\}\), where \(\theta\neq 2n\pi\), \(n\in\mathbb{Z}\),  such that \(\measuredangle(\mv g_{sp}^H\mv w_p^\prime)=\measuredangle(\mv g_{EAP,p}^H\mv v_p^\ast)\). Hence, it follows that
\begin{align}
\vert\mv g_{sp}^H\mv w_p^\prime+\mv g_{EAP,p}^H\mv v_p^\ast\vert&=\vert(\vert\mv g_{sp}^H\mv w_p^\prime\vert+\vert\mv g_{EAP,p}^H\mv v_p^\ast\vert)\exp\{j\measuredangle(\mv g_{EAP,p}^H\mv v_p^\ast)\}\vert\notag\\
&=\vert\mv g_{sp}^H\mv w_p^\ast\vert+\vert\mv g_{EAP,p}^H\mv v_p^\ast\vert\notag\\
&\stackrel{\rm (a)}{>}\vert\mv g_{sp}^H\mv w_p^\ast+\mv g_{EAP,p}^H\mv v_p^\ast\vert\notag\\
&\ge \sqrt{t^\ast(\mv g_{sp}^H\mv Q_s^\ast\mv g_{sp}+\sigma_{PR}^2)}, \label{eq:align w_p and v_p}
\end{align}
where ``\(\ge\)'' in \({\rm (a)}\) holds strictly, as a result of \(\measuredangle(\mv g_{sp}^H\mv w_p^\ast)\neq\measuredangle(\mv g_{EAP,p}^H\mv v_p^\ast)\). According to \eqref{eq:align w_p and v_p}, it holds true that \(\exists \mv w_p^{\prime\prime}=\delta\mv w_p^\prime\), where \(\delta\in[0,1)\), such that
\begin{align}
\vert\mv g_{sp}^H\mv w_p^\prime+\mv g_{EAP,p}^H\mv v_p^\ast\vert>\vert\mv g_{sp}^H\mv w_p^{\prime\prime}+\mv g_{EAP,p}^H\mv v_p^\ast\vert
=&\delta\vert\mv g_{sp}^H\mv w_p^\ast\vert+\vert\mv g_{EAP,p}^H\mv v_p^\ast\vert\notag\\
>&t^\ast(\mv g_{sp}^H\mv Q_s^\ast\mv g_{sp}+\sigma_{PR}^2).
\end{align}
Meanwhile, we change the solution of \(\varrho^\ast\) to be \(\varrho^\prime\) which is expressed as
\begin{align}
\varrho^\prime=\frac{{\sf Tr}(\mv Q_s^\ast)+\delta^2\|\mv w_p^\ast\|^2}{{\sf Tr}(\mv Q_s^\ast)+\|\mv w_p^\ast\|^2}\varrho^\ast<\varrho^\ast.
\end{align}
So far, by changing the solution from \(\varrho^\ast\) to  \(\varrho^\prime\), it is observed that the constraints \eqref{eq:SINR of the ST superior to that of the EAPs} and  \eqref{eq:SINR of the ST larger than t} still hold, for the fact that the LHS of \eqref{eq:SINR of the ST superior to that of the EAPs} is a monotonically decreasing function over \(\varrho\in[0,1]\). Then we take the next step of changing \(t^\ast\) to \(t^\prime\) as follows.
\begin{align}
t^\prime=\min\left\{(1-\varrho^\prime)P_p\mv h_{p,ST}^H\mv A^{-1}(\varrho^\prime,\mv X^\ast)\mv h_{p,ST},\frac{\left\vert\mv g_{sp}^H\mv w_p^{\prime\prime}+\mv g_{EAP,p}^H\mv v_p^\ast\right\vert^2}{\mv g_{sp}^H\mv Q_s^\ast\mv g_{sp}+\sigma_{PR}^2}\right\}>t^\ast.
\end{align}

Consequently, by changing the solution of \(\mv w_p^\ast\), \(\varrho^\ast\) and \(t^\ast\) to \(\mv w_p^{\prime\prime}\), \(\varrho^\prime\), and \(t^\prime\) without changing others, we find that a larger objective value to \(\mathrm{(P1.1)}\) is achieved without violating any other constraints, due to the increasing \(\log_2(1+t^\prime)\). This nevertheless contradicts to the claimed optimality achieved by \(t^\ast\). Hence, the proof is complete.

\section{Proof of Proposition~\ref{prop:convergence and KKT}}\label{appendix:proof of convergence and KKT}
Algorithm~\ref{alg:Algorithm I} will generate a sequence of feasible points $ \{ {\mv X}^{(n)}, {\mv Q}_s^{(n)}, {\mv w}_p^{(n)}, {\mv v}_p^{(n)}, t^{(n)}\} $ to problem $ \mathrm{(P1.1\text{-}1)} $, since for each iteration the feasible region to \(\mathrm{(P1.1\text{-}1^\prime)}\) is always a subset of that to $ \mathrm{(P1.1\text{-}1)} $. For the $n$th iteration, \(\{\bar{\mv X}^{(n)}, \bar{\mv Q}_s^{(n)}, \bar{\mv w}_p^{(n)}, \bar{\mv v}_p^{(n)}, \bar t^{(n)}\}\) is chosen as a feasible solution to \(\mathrm{(P1.1\text{-}1^\prime)}\), while \(\{ {\mv X}^{(n+1)}, {\mv Q}_s^{(n+1)}, {\mv w}_p^{(n+1)}, {\mv v}_p^{(n+1)}, t^{(n+1)}\} $ is the returned optimal solution to \(\mathrm{(P1.1\text{-}1^\prime)}\). Hence, denoting the objective function of \(\mathrm{(P1.1\text{-}1^\prime)}\) as \(h_1\) with its (implicit) dependence on the optimization variables omitted,  $h_1^{(n+1)} \geq h_1( \bar{\mv X}^{(n)}, \bar{\mv Q}_s^{(n)}, \bar{\mv w}_p^{(n)}, \bar{\mv v}_p^{(n)}, \bar t^{(n)} )=h_1^{(n)}$. As a result, we arrive at a non-decreasing sequence $ \{ h_1^{(n)} \} $.

Furthermore, we show that the solutions generated by the sequence $ \{ {\mv X}^{(n)}, {\mv Q}_s^{(n)}, {\mv w}_p^{(n)}, {\mv v}_p^{(n)}, t^{(n)}\} $ are bounded. In fact, the boundedness for $ {\mv X}^{(n)} $ and $ {\mv v}_p^{(n)} $ can be justified via constraints \eqref{eq:per-EAP power constraint for WIT} and \eqref{eq:per-EAP power constraint for WPT in terms of X}, respectively. Consequently, it holds true that the nuclear norm of \(\mv Q_s^{(n)}\) and the \(\ell^2\)-norm of \(\mv w_p^{(n)}\), which are bounded by $ \eta\varrho P_{\rm EH}({\mv X}^{(n)}) $, is also bounded from the above. As for $ t^{(n)} $, according to \eqref{eq:SINR of the ST larger than t with fixed rho}, it implies that
\begin{align}
t^{(n)} \le &(1-\bar\varrho)P_p\mv h_{p,ST}^H{{\mv A}^{(n)}}^{-1}\mv h_{p,ST} \notag\\
\le & (1-\bar\varrho)P_p\lambda_{\max}({{\mv A}^{(n)}}^{-1})\|\mv h_{p,ST}\|^2 \notag\\
\le &\Myfrac{(1-\bar\varrho)P_p\|\mv h_{p,ST}\|^2}{(1-\bar\varrho)\sigma_{n_a}^2+\sigma_{n_c}^2}
\end{align} where \(\lambda(\cdot)\) denotes the eigenvalue of the associated matrix.

Owing to the continuity of $ h_1 $,  it follows that $ \{h_1^{(n)}\} $ is also bounded from the above. Hence, the non-decreasing sequence $ \{h_1^{(n)}\} $ is convergent to a real number denoted by $ h_1^\ast $. Furthermore, according to Bolzano-Weierstrass theorem, the bounded sequence $ \{ {\mv X}^{(n)}, {\mv Q}_s^{(n)}, {\mv w}_p^{(n)}, {\mv v}_p^{(n)}, t^{(n)}\} $ has at least one convergent subsequence, the accumulation point of which is denoted by $ \{ {\mv X}^\ast, {\mv Q}_s^\ast, {\mv w}_p^\ast,$ ${\mv v}_p^\ast, t^\ast\} $. Therefore, we automatically arrive at \(h_1({\mv X}^\ast, {\mv Q}_s^\ast, {\mv w}_p^\ast,\) \({\mv v}_p^\ast, t^\ast)=h_1^\ast\). A KKT point of problem $ \mathrm{(P1.1\text{-}1)} $, namely, $ \{ {\mv X}^\ast, {\mv Q}_s^\ast, {\mv w}_p^\ast,\) \({\mv v}_p^\ast, t^\ast\} $ is thus obtained \cite[Theorem 1]{marks1978technical}.

A similar proof can be applied to problem \(\mathrm{(P1.2\text{-}1)}\) and is thus omitted here for brevity.


\bibliographystyle{IEEEtran}
\bibliography{CRN_ref}
\end{document}